\documentclass[%
 preprint,
showpacs,showkeys,preprintnumbers,longbibliography,
 amsmath,amssymb,
 aps,
 pra,
]{revtex4-1}

\usepackage{color}
\usepackage{amsmath}
\usepackage{graphicx}
\usepackage{dcolumn}
\usepackage{bm}
\usepackage{subcaption}
\usepackage{makecell}
\DeclareMathOperator\erf{erf}

\begin{document}


\title{On the velocity distribution function of spontaneously evaporating atoms}

 \author{Sergiu Busuioc}
  \altaffiliation{School of Engineering, The University of Edinburgh, Edinburgh, EH9 3FB, United Kingdom}
 \author{Livio Gibelli}
  \altaffiliation{School of Engineering, The University of Edinburgh, Edinburgh, EH9 3FB, United Kingdom}
 \author{Duncan A. Lockerby}
  \altaffiliation{School of Engineering, University of Warwick, Coventry CV4 7AL, United Kingdom}
 \author{James E. Sprittles}
  \altaffiliation{Mathematics Institute, University of Warwick, Coventry CV4 7AL, United Kingdom}
%
%

\date{\today}

\begin{abstract}
Numerical solutions of the Enskog-Vlasov (EV) equation are used to determine the velocity distribution function of atoms spontaneously evaporating into near-vacuum conditions.
It is found that an accurate approximation is provided by a half-Maxwellian including a drift velocity combined with different characteristic temperatures for the velocity components normal and parallel to the liquid-vapor interface. The drift velocity and the temperature anisotropy reduce as the liquid bulk temperature decreases but persist for relatively low temperatures corresponding to a vapor behaviour which is only slightly non-ideal.
Deviations from the undrifted isotropic half-Maxwellian are shown to be consequences of collisions in the liquid-vapor interface which preferentially backscatter atoms with lower normal-velocity component.
\end{abstract}

\maketitle


\section{Introduction}

Evaporation and condensation are ubiquitous processes which play crucial roles in different areas, from physics and chemistry to biology. 
The fundamental understanding of the heat and mass transfer across the liquid-vapor interface is not only of theoretical interest but also of considerable relevance to a wide range of applications. 
Examples include the distillation process in high vacuum conditions~\cite{ST13,LX14}, the heating and evaporation of fuel droplets~\cite{S17}, the thermal management of modern electronic devices~\cite{JESGEL19}, and the flash boiling in gasoline direct-injection sprays~\cite{CLPP20}.

The modelling of fluid flows with phase change is challenging since multiple characteristic time and length scales are involved.
The liquid and vapor bulk phases are well described at the macroscopic scale by hydrodynamic equations, while the separating region has a more complex structure, with sharp gradients in flow variables which manifest as jumps at the macroscopic scale. 
In non-equilibrium conditions, the region between the two bulk phases comprises two sub-regions, namely the liquid-vapor interface, whose  characteristic length is of the order of the molecular diameter~\cite{RW13}, and the tipically larger Knudsen layer that extends a few mean free paths into the vapor phase~\cite{S00}.

The standard kinetic theory studies on evaporation and condensation processes focus on the vapor dynamics in the Knudsen layer next to the liquid phase.
The liquid-vapor interface is simplified to a structureless surface bounding the vapor and the molecular exchanges with the liquid phase are dealt with using a phenomenological boundary condition. More specifically, the molecular flux coming out of the liquid-vapor surface is assumed to have two contributions: atoms spontaneously leaving the liquid bulk (evaporated atoms) and atoms which are backscattered into the vapor after impinging on the liquid-vapor interface (reflected atoms). 
The distribution function of evaporating atoms is usually described by a half-Maxwellian while the Maxwell's scattering kernel is the most widely used choice to describe the molecular reflection from the liquid phase.
Two phenomenological coefficients are also introduced to weight the relative importance of the evaporation and condensation fluxes.

In spite of the increasing number of experimental, theoretical and numerical studies, the full understanding of evaporation processes is still lacking and numerous issues remain unresolved. As an example, while the evaporation and condensation coefficients are generally accepted to be lower than unity, experimental measurements for water have provided values that span three orders of magnitude~\cite{PW16}. Furthermore, a temperature discontinuity has been experimentally found across the water liquid-vapor interface that is much larger in magnitude and in the opposite direction to that predicted by classical kinetic theory or nonequilibrium thermodynamics~\cite{fang1999temperature}.
	 

A lot of effort has been thus expended in assessing the physical appropriateness of kinetic boundary conditions at the liquid-vapor interface mostly through molecular dynamics (MD) simulations and mean-field kinetic approximations of simple liquids.
These studies can be roughly grouped into two categories depending on whether the liquid evaporation occurs into vapor~\cite{C36,BL19,TTM99,MFYH04,XSC11,XCS12,KKW14,KSKFW17} or into near vacuum conditions~\cite{ZA97,IYF04, FGL05, FGLS18, HV19}. In this latter process the backscattered vapor component is virtually absent and, therefore, the distribution function of evaporated atoms is evaluated without any ambiguity.

		
Accurate MD simulations of liquid argon evaporating into vapor have shown that atoms leaving the liquid-vapor interface are distributed with excellent approximation according to a half-Maxwellian while, separately, evaporated and reflected atoms are not~\cite{TTM99,MFYH04}.
More specifically, the velocity distribution function of evaporated atoms shows a fatter tail compared to a Maxwellian while the one of reflected atoms has a reduced tail and larger density in the low velocity region. However, deviations are described differently, namely through the dependency of the condensation/evaporation coefficient on the normal-velocity component of atoms impinging on the liquid surface~\cite{TTM99} or by a drifted half-Maxwellian~\cite{MFYH04}.

Using MD simulations, deviations from a half-Maxwellian have also been revealed in the case of n-dodecane, i.e. more complex molecules, but the distribution function of evaporated atoms is fitted by an anisotropic half-Maxwellian (also referred to as bi-Maxwellian) with the temperature normal to the liquid-vapor interface larger than the parallel one~\cite{XSC11,XCS12}.
The hypothesis that an anisotropic half-Maxwellian provides a better approximation of the distribution function close to the evaporating surface was conjectured for the first time in Ref.~\cite{C36} and the theoretical analysis has been more recently extended to include the case of an evaporation coefficient less than one~\cite{BL19}.




According to review above, most of the MD studies of liquids evaporating into vapor agree that the distribution function of evaporated atoms deviates from the half-Maxwellian, albeit there is no consensus on the fitting function. The simulation results of liquids evaporating into near vacuum conditions are even less conclusive in that deviations have not always been found.

In some MD studies, it has been shown that, in the limit of a low-density vapor, evaporated argon atoms are distributed according to a half-Maxwellian~\cite{ZA97, ZKLSA19}. This conclusion has been explained by the fact that atoms leave the condensed phase due to collisions from the edge of the liquid-vapor interface where the binding energy is negligible, whence the half-Maxwellian shape.

In other MD studies, the velocity distribution function of spontaneously evaporating atoms has been determined to be nearly a half-Maxwellian at low temperatures, but a tail fatter than the one of a half-Maxwellian has been found normal to the liquid-vapor interface at high temperatures~\cite{IYF04,HV19}.




The validity of standard kinetic boundary conditions at the liquid-vapor interface has also been assessed based on the Enskog-Vlasov (EV) equation.
This kinetic equation has the capability of describing both the liquid and vapor phases, including the interface region and has many attractive features. Compare to MD simulations, its numerical solution requires less computational effort and, unlike macroscopic approaches, such as diffuse interface models and sharp interface methods, it can capture the nonequilibrium effects in the liquid-vapor interface and in the Knudsen layer.

The EV equation has been used to study the evaporation into near vacuum of a thin liquid film~\cite{FGL05}, the steady evaporation/condensation flow between two planar liquid films kept at different temperatures~\cite{KKW14}, and, more recently, the evaporation of multi-component substances into vapor and vacuum~\cite{KSKFW17,FGLS18}.

Remarkably, in spite of the simplified treatment of pair correlations in the dense phase, the results obtained by the EV equation closely resemble the ones provide by MD simulations. More specifically, it is generally concluded that evaporated atoms are distributed according to a half-Maxwellian with liquid bulk temperature even though a slight anisotropy between the velocity components normal and parallel to the liquid-vapor interface has been found at high evaporation temperatures. 

The foregoing literature review shows that uncertainties remain on the statistical features of atoms spontaneously emitted from the liquid phase. Most of the studies observed that evaporated atoms are not distributed according to a half-Maxwellian but none has systematically evaluated the deviations as a function of the evaporation temperature. Furthermore, the physical mechanism which leads the velocity distribution function of evaporated atoms to differ from a half-Maxwellian has not been convincingly explained.

The aim of the present work is thus twofold. 
First, to accurately determine the distribution function of atoms evaporating into near-vacuum conditions by numerically solving the EV equation. 
Compared to MD simulations, the numerical solution of this kinetic equation requires less computational effort and, therefore, permits one to get results with the required level of accuracy for the analysis undertaken in this study~\cite{FB17}. 
Second, to provide a convincing explanation of the deviations from the half-Maxwellian.  

The rest of the paper is organised as follows. 
Section~\ref{sec:formulation} outlines the mathematical formulation of the evaporation of a monatomic single-component liquid into near-vacuum condition based on the EV equation and outlines the particle method of solution.
Section~\ref{sec:results} contains a brief description of the computational setup and presents the main results of the paper. In particular, it is shown that the reduced distribution function of spontaneously evaporated atoms are well approximated by a drifted anisotropic half-Maxwellian. Furthermore, it is shown, by numerical evidence and a simple mathematical model, that deviations from an isotropic half-Maxwellian are due to the atoms' collisions in the liquid-vapor interface.
Section~\ref{sec:conclusions} summarises and comments on the main findings of the paper.

\section{Mean-field kinetic theory approach}
\label{sec:formulation}
\subsection{The Enskog-Vlasov equation}

Let us consider a fluid composed of spherical and identical atoms of mass $m$ and diameter $a$ interacting through the Sutherland potential given by a superposition of a hard sphere potential and an attractive soft potential tail:
\begin{equation}
 \phi(\rho)=
 \begin{cases}
   +\infty \,,\quad   &\rho < a, \\
   \displaystyle{-\phi_a\left( \frac{\rho}{a}\right)^{-\gamma}}\, ,\quad &\rho\geq a,
  \end{cases}
\end{equation}
where $\rho=||\bm{r}_1-\bm{r}||$ is the distance between the atoms at position $\bm{r}_1$ and $\bm{r}$ while the two positive constants $\phi_a$ and $\gamma$ 
define the depth of the potential well and the range of the soft interaction, respectively.
Karkheck et al \cite{KS81} derived the {\it exact} evolution equation for the one-particle distribution function of this system of atoms, $f(\bm{r},\bm{v},t)$, but this equation is of little use since it involves the two-particle distribution function. In order to obtain a closed equation, two simplifying assumptions are introduced in the particle dynamics, namely long-range particle correlations are neglected while short-range particle correlations are approximated by following the Enskog theory originally derived for dense gases.
By adopting these assumptions, the following closed equation for the evolution of $f(\bm{r},\bm{v},t)$ is obtained:
\begin{subequations}
 \label{eq:ev}
 \begin{equation}
 \label{eq:ev_general}
  \frac{\partial f}{\partial t}
  +\bm{v}\cdot\nabla_{\bm{r}}f+\frac{\mathcal{F}[n]}{m} \cdot \nabla_{\bm{v}}f=\mathcal{C}_E[f],
 \end{equation}
where square brackets denote functional dependence. 
In Eq.~\eqref{eq:ev_general}, $\mathcal{F}(\bm{r},t)$ is the self-consistent force field generated by the soft attractive tail which reads:
\begin{equation}
 \mathcal{F}[n]=\int_{||\bm{r}_1-\bm{r}||>a} \frac{d\phi_a(\rho
 )}{d\rho}\frac{\bm{r}_1-\bm{r}}{||\bm{r}_1-\bm{r}||}n(\bm{r_1})d\bm{r_1},
 \end{equation}
and the hard-sphere collision integral $\mathcal{C}_E(f,f)$ is given by:
 \begin{multline}
 \mathcal{C}_E[f]=
a^2\int(\bm{v}_r\cdot\bm{\hat{k}})^+d\bm{v}_1d^2\bm{\hat{k}}
\left\{\chi\left[n\left(\bm{r}+\frac{a}{2}\bm{\hat{k}},t\right)\right] f(\bm{r}+a\bm{\hat{k}},\bm{v}_1^*,t) f(\bm{r},\bm{v}^*,t)-\right.\\
  \left.\chi\left[n\left(\bm{r}-\frac{a}{2}\bm{\hat{k}},t\right)\right] f(\bm{r}-a\bm{\hat{k}},\bm{v}_1,t) f(\bm{r},\bm{v},t)  \right\},\label{eq:enskog}
\end{multline}
where $(\cdot)^{+}$ indicates that the surface integral is restricted to the half-sphere for which $\bm{v}_r\cdot\bm{\hat{k}}>0$ and $\chi[n]$ is the contact value of the pair correlation function in a hard-sphere fluid in equilibrium with number density $n$.
\end{subequations}
In the Standard Enskog Theory (SET), $\chi$ is approximated by using the value of the pair correlation function in a fluid in uniform equilibrium with the value of the density at the contact point of the two colliding atoms. An approximate, but accurate expression for $\chi_{\mbox{\tiny SET}}$ can be obtained from the equation of state of the hard-sphere fluid proposed by Carnahan and Starling~\cite{CS69}, as:

\begin{equation}
\chi_{\mbox{\tiny SET}}(n)=\frac{1}{nb}\left(\frac{p^{hs}}{n k_B T}-1\right)=\frac{1}{2}\frac{2-\eta}{(1-\eta)^3}, 
\qquad b=\frac{2\pi a^3}{3},\qquad \eta=\frac{\pi a^3 n}{6}
\label{eq:carnstar}.
\end{equation}
where $p^{hs}$ is the pressure of a system of hard spheres and $k_B$ is the Boltzmann constant.
In the present work, following the Fischer-Methfessel approach~\cite{FM80}, the actual value of the density at the contact point is replaced with the value of the density field averaged over a spherical volume of radius $a$, namely:
\begin{subequations}
\begin{equation}
\chi[n](\bm{r},\bm{r}-a\bm{\hat{k}})=\chi_{\mbox{\tiny SET}}\left(\overline{n}\left(\bm{r}-a \frac{\bm{\hat{k}}}{2}\right)\right),
\end{equation}
where
\begin{equation}
\overline{n}(\bm{r},t)=\frac{3}{4\pi a^3}\int_{\mathbb{R}^3} n(\bm{r}_*,t)w(\bm{r},\bm{r}_*)\,d\bm{r}_*, \hspace{1cm}
w(\bm{r},\bm{r}_*)=\left\{
\begin{array}{cc} 
1, &\qquad \|\bm{r}_*-\bm{r}\|<a, \\ 
0, & \qquad \|\bm{r}_*-\bm{r}\|>a.
\end{array}
\right.
\end{equation}
\end{subequations}
The kinetic equation~\eqref{eq:ev} is usually referred to as the Enskog-Vlasov (EV) equation~\cite{S67,G71,KS81,FGL05,BB19}.
This kinetic equation has been applied to investigate a wide range of two-phase flows, including condensation/evaporation processes~\cite{FGL05,KKW14,FBG19}, liquid menisci in nano-channels~\cite{BFG15}, and its mathematical properties have been extensively studied in connection with the liquid-vapor transition~\cite{TMHH18,BB18}.  

\subsection{Particle method of solution}
 
In this work, the EV equation is solved numerically by an extension of the original Direct Simulation Monte-Carlo (DSMC) scheme to dense fluids \cite{F97b}. A thorough description of the numerical scheme and the analysis of its computational complexity is given in Ref.~\cite{FBG19}. 

For EV simulations, the main framework of DSMC scheme used to solve the Boltzmann equation is preserved, with modifications occurring in the collision algorithm due to the nonlocal structure of the Enskog collision operator.
The distribution function is represented by $N$ computational particles:
\begin{equation}
f(\bm{r},\bm{v},t)= \sum_{i=1}^{N} \delta{\left(\bm{r}-\bm{r}_i(t)\right)} \delta(\bm{v}-\bm{v}_i(t)),
\end{equation}
where $\bm{r}_i$ and $\bm{v}_i$ are the position and the velocity of the $i$th particle at time $t$, respectively.

The distribution function is updated by a fractional-step method based on the time-splitting of the evolution operator in two sub-steps, namely free streaming and collision.
In the first stage, the collisions between particles are neglected and the distribution function is advanced from $t$ to $t+\Delta t$ by solving the equation:
\begin{equation}
\label{eq:stage_I}
   \frac{\partial f}{\partial t}
  +\bm{\bm{v}}\cdot\nabla_{\bm{r}}f+\frac{\mathcal{F}[n]}{m}\cdot\nabla_{\bm{\bm{v}}}f=0,
\end{equation}
which translates into updating the positions and velocities of the computational particles according to:
\begin{subequations}
\begin{align}
 \bm{r}_i(t+\Delta t)&=\bm{r}_i(t)+\bm{v}_i\Delta t+\frac{\mathcal{F}[n(t)]}{m}\frac{(\Delta t)^2}{2},\\
\bm{v}_i(t+\Delta t)&=\bm{v}_i(t)+\frac{\mathcal{F}[n(t)]}{m}\Delta t.
 \end{align}
\end{subequations}
In the second stage, the short range hard-sphere interactions are considered and the updating rule is given by:
\begin{equation}
 f(\bm{r},\bm{v},t+\Delta t)= \tilde{f}(\bm{r},\bm{v},t+\Delta t)+\mathcal{C}_E[\tilde{f}]\Delta t.
\end{equation}
During this stage, particles' positions $\bm{x}_i$ are unchanged while their velocities $\bm{v}_i$ are modified according to stochastic rules which essentially correspond to the Monte Carlo evaluation of the collision integral given by Eq.~\eqref{eq:enskog}.

The macroscopic quantities are obtained by time averaging the particles' microscopic states. Note that for steady flows simulations, as the ones considered below, the averaging time can be long enough to obtain accurate results without the need of using a large number of computational particles.

\section{Evaporation of monatomic liquid into near vacuum}
\label{sec:results}

\begin{figure}
\includegraphics[width=0.65\linewidth]{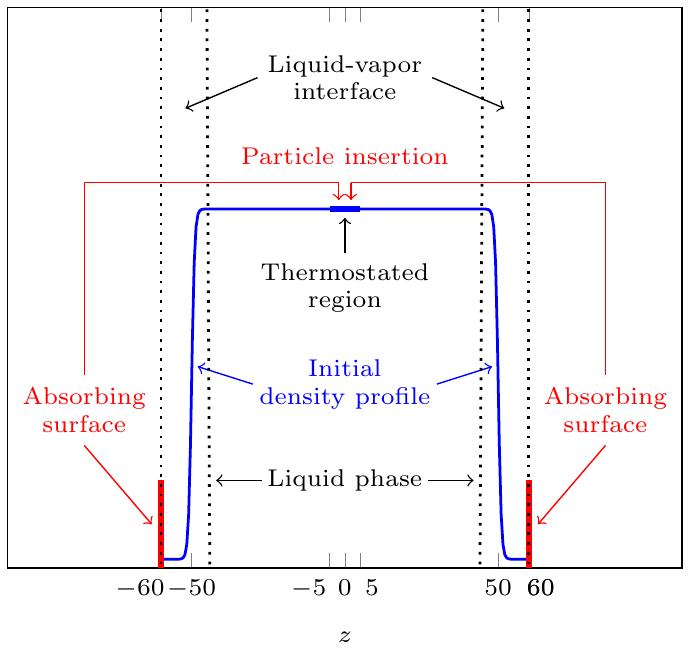}
\caption{Schematic of the computational setup. The simulation domain is a finite symmetric interval with perfectly absorbing boundary conditions at $z=\pm 60a$. The initial condition consists of a liquid slab in equilibrium with its vapor placed in the middle of the simulation domain.}
\label{fig:schematic}
\end{figure}

\subsection{Computational setup}
\label{sec:computational_setup}
The schematic of the computational setup is shown in Fig.~\ref{fig:schematic}.
The simulation domain is a finite symmetric interval $z\in[-L,L]$ with $L=60a$ and homogeneous along the $x$ and $y$ axis. Perfectly absorbing boundary conditions are assumed at $z=\pm L$, namely computational particles reaching the boundaries are removed from the simulation domain. Initially, a liquid slab in equilibrium with its vapor is considered in the middle of the simulation domain. This initial state has been obtained by placing a liquid slab in $z\in[-L_s,L_s]$, with $L_s=50a$, gas elsewhere, and using periodic boundary conditions at $z=\pm L$. Afterwards, the system has been let to evolve until the equilibrium steady state is reached.
The distance between the liquid-vapor interfaces and the absorbing surfaces is about $10a$, namely large enough not to influence the interface but, at the same time, sufficiently small to minimise the backscattered flow.

The system is clearly symmetric with respect to the $z$ axis and, therefore, in principle, the computational setup might be simplified by considering only half of the domain. However, this would lead to the need of imposing a specular boundary condition at the center of the liquid slab which, among other things, would require a tricky treatment of the mean force field. Furthermore, results in the two halves of the domain can be superimposed and, therefore, the larger computational effort of simulating all the system is used to reduce the statistical noise.

The simulation domain is divided into 2400 cells with size $\Delta z=a/20$ and the time step is $\Delta t=2\times 10^{-4}\,a/(R T_0)^{1/2}$, where $R=k_B/m$ is the specific gas constant and $T_0$ is the reference temperature.
The number of computational particles is set to $1.2\times10^6$ and
made equal to the number of real atoms by a proper choice of the cross section normal to the non-homogeneous direction $z$.
The interaction parameters were chosen to be $\phi_a/(k_B T_0)=1$ and $\gamma=6$ so as to match the same far field behavior as the 12-6 Lennard-Jones potential~\cite{HBC64}. 
The study of evaporation of a liquid slab is carried out at the following temperature values $T_\ell/T_c=\{0.53,0.596,0.663,0.729\}$, where $T_c$ is the critical temperature $T_c/T_0=0.754632$ as a consequence of the interaction parameters chosen~\cite{FGL05}.
In this temperature range, the vapor phase contains a number of particles sufficient to limit the statistical noise of the results (which are obtained by averaging particles' properties) but, at the same time, is dilute enough to behave as an ideal gas. The Andersen thermostat~\cite{A80} is applied in the central part of the liquid slab, $10a$ wide, to prevent the progressive cooling of the system and keep its temperature to the constant value of $T_\ell$. This thermostat is chosen for its simplicity and computational efficiency. 

Note that, in principle, during the evaporation into vacuum, the liquid slab is consumed and the liquid-vapor interfaces slowly recede from the absorbing surfaces. The flow is thus an unsteady process and the possibility to evaluate the macroscopic quantities by time-averaging is jeopardized.
In order to circumvent this difficulty, in the present work the evaporation process is studied in a frame of reference fixed relative to the liquid-vapor interface using the following procedure.
The simulation advances until the number of particles reaching the absorbing surfaces equals or exceeds $2N_\ell$, where $N_\ell$ is the average number of particles per cell in the liquid bulk at temperature $T_\ell$. When this happens, the simulation stops. All particles are then moved by $\Delta z$ towards the closest absorbing surface and the empty gap, which forms in the centre of the simulation domain, is filled with $2N_\ell$ particles sampled from a Maxwellian with temperature $T_\ell$. Afterwards, the simulation is restarted. Note that the interface movement is negligibly small during two successive applications of this procedure. In this steady evaporation framework, simulations are run for the time duration $2000\,a/(RT_0)^{1/2}$ (equivalent to $10^7$ iterations).

\subsection{Simulation results}

\subsubsection{Macroscopic quantities}
\label{sec:macro}

\begin{figure}
 	\begin{subfigure}[b]{0.49\textwidth}
		\begin{center}
			\includegraphics[width=\linewidth]{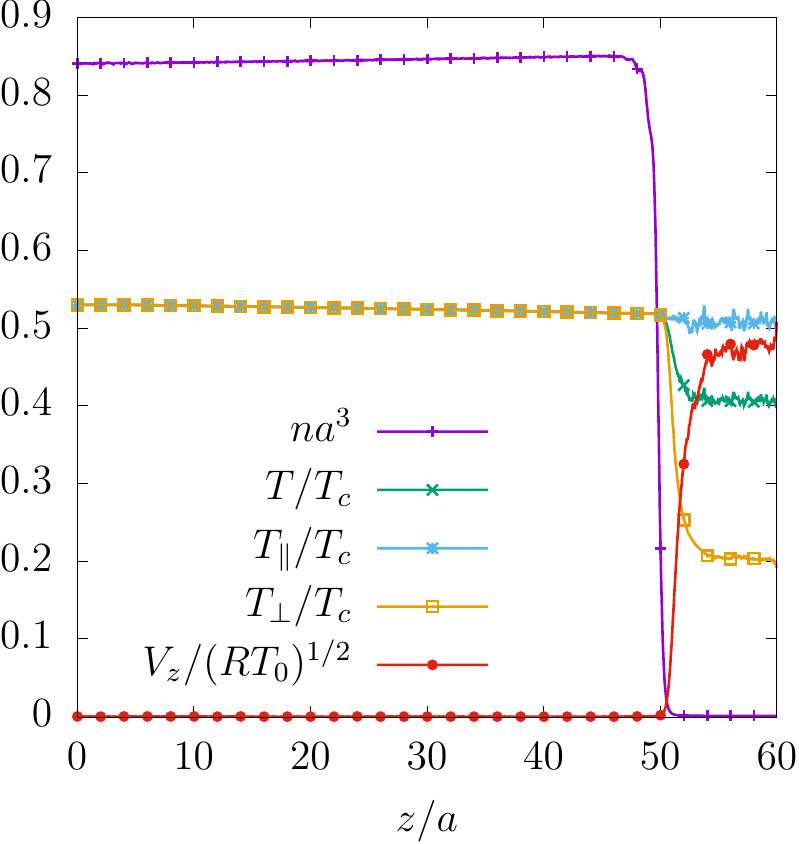}
			\subcaption{$T_\ell/T_c=0.530$}
		\end{center} 
	\end{subfigure} \hfill
	\begin{subfigure}[b]{0.49\textwidth}
		\begin{center}
			\includegraphics[width=\linewidth]{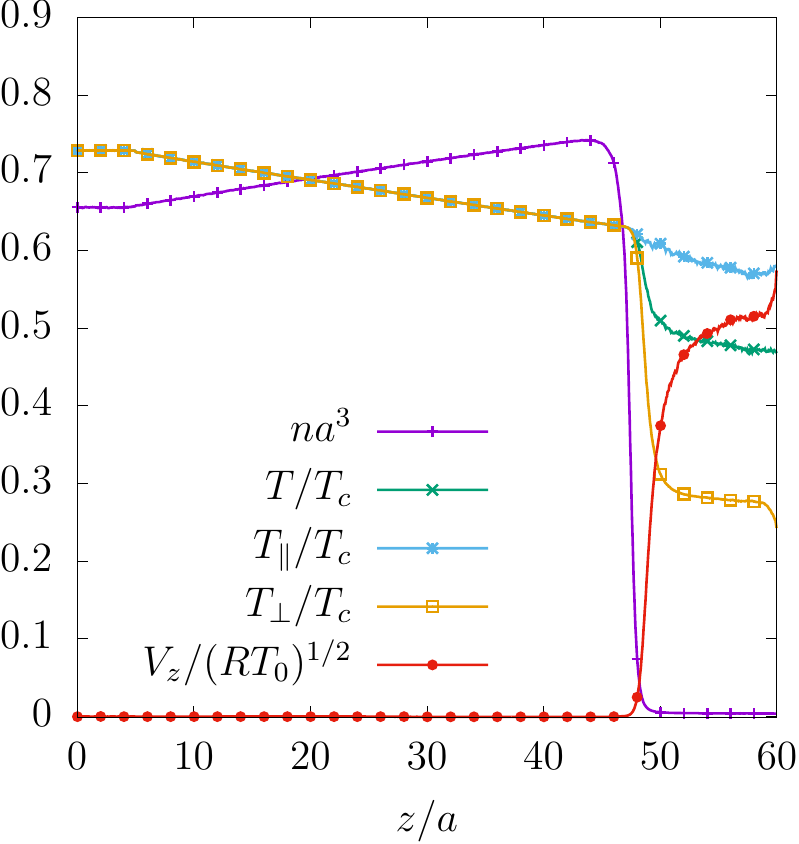}
			\subcaption{$T_\ell/T_c=0.729$}
		\end{center}		
	\end{subfigure}
		\caption{Dimensionless density, $na^3$, mean velocity in the $z$-direction, $V_z/(RT_0)^{1/2}$, normal, $T_\perp/T_c$, parallel, $T_\parallel/T_c$, and total, $T/T_c$, temperatures for (a) the lowest and (b) the highest liquid bulk temperatures considered in the simulation campaign. 
		\label{fig:macro_all}}
\end{figure}

A broad picture of a liquid slab evaporating into near-vacuum is provided by observing the behaviour of the macroscopic quantities across the the liquid and vapor phases shown in Fig.~\ref{fig:macro_all} for the lowest ($T_\ell/T_c=0.53$) and the highest ($T_\ell/T_c=0.729$) temperatures considered. Due to the symmetry of the system, only half of the simulation domain is shown.

The minimum density in the liquid phase is found in the central part of the domain where the liquid is thermostated, while the maximum is reached close to the liquid-vapor interface, with an almost linear profile in between them. This behavior is due to the evaporation cooling of the liquid slab, which causes a temperature decrease in the liquid region which is not thermostated.
After the drop in the liquid-vapor interface, the temperature in the vapor phase levels off but still exhibits a small gradient. This clearly indicates that some collisions between atoms occur in the gap region between the edge of the interface at the vapor side and the absorbing surface.

The normal and parallel temperatures, which are defined based on the velocity components normal and parallel to the liquid-vapor interface, superimpose up to a point located at about half of the interface width. Their separation indicates that the local thermodynamic quasi-equilibrium condition breaks down, marking the beginning of a transition layer which extends a few molecular diameters into the low density region. In the following, we refer to the point at which the separation of temperatures occurs as the `separation point', denoted $z_s$, and the corresponding value of the temperature as the `separation temperature', denoted $T_s$. 
After the separation point the parallel temperature profile follows a similar slope as the one in the liquid bulk, while the normal temperature exhibits a sharp drop coupled with an increase in the mean velocity component normal to the liquid-vapor interface.

Note that the normal temperature and the mean velocity in the $z$-component show gradients in a thin layer close to the boundary of the computational domain, especially for the highest temperature case shown in Fig.~\ref{fig:macro_all}(b). This behaviour is not a numerical artifact but it is due to the the discontinuity of the density field at the absorbing surface, which results in a mean force field directed towards the liquid acting on the atoms in this boundary region. 

\subsubsection{Velocity distribution function of evaporated atoms}
\label{sec:vdf}

\begin{table}[t!]
	\begin{center}
		\begin{tabular*}{\columnwidth}{@{\extracolsep{\stretch{1}}}*{5}{lcccc}@{}}
			\toprule
			    & Parameters & {Undrifted anisotropic} & {Drifted isotropic}& {Drifted anisotropic} \\ \hline
			$T_\ell/T_c=0.530$ & $\xi/(RT_0)^{1/2}$ & --- & $0.026366\pm0.00388$ & $0.03427\pm0.01364$ \\
			&	$\theta_\perp/T_c$ & $0.5278\pm0.00254$ & $0.51238\pm0.00176$ & $0.5072\pm0.00852$   \\
			&	$\theta_\parallel/T_c$ & $0.51238\pm0.00176$ & $0.51238\pm0.00176$ & $0.51238\pm0.00176$   \\
			&	$\chi^2$ &$0.03016$ &$0.02866$ &$0.02856$\\
			\hline
			$T_\ell/T_c=0.729$ & $\xi/(RT_0)^{1/2}$ & --- & $0.10903\pm0.00246$ & $0.1628\pm0.0056$\\
			& $\theta_\perp/T_c$ &$0.66987\pm0.00348$&$0.59937\pm0.00087$ & $0.55924\pm0.00397$ \\ 
			&	$\theta_\parallel/T_c$ & $0.59937\pm0.00087$ & $0.59937\pm0.00087$ & $0.59937\pm0.00087$   \\
			& $\chi^2$ &$0.03911$&$0.01163$&$0.00688$   \\
			\hline \hline	
		\end{tabular*}
	\end{center}
	\caption{Estimates of the fitting parameters, $\xi,\, \theta_\perp$, $\theta_\parallel$ alongside the asymptotic standard error, and the residual sum of squares, $\chi^2$, of the three considered fitting velocity distribution functions for the lowest and the highest liquid bulk temperatures considered in the simulation campaign.}
	\label{tab:fits}
\end{table}

The statistical features of spontaneously evaporated atoms have not been systematically assessed until now. According to some studies their velocity distribution function can be approximated by an anisotropic half-Maxwellian~\cite{CLPG11,IYF04,FGLS18} while others pointed out the presence of a velocity drift~\cite{MFYH04}.
Here, the velocity distribution function of evaporated atoms is evaluated based on the particles collected at the absorbing surfaces. As specified in Sec.~\ref{sec:computational_setup}, these surfaces are placed at the edge of the liquid-vapor interface where kinetic boundary conditions can be formulated. Three different Maxwellian-like functional forms are tested for fitting:
\begin{subequations}
	\begin{align}
	\label{eq:I}
	\hspace{-0.36cm} \mbox{Undrifted Anisotropic:}\hspace{0.1cm} & f(v_\parallel,v_\perp) = \frac{C_1}{(2\pi R)^{3/2}\theta_\perp\theta_\parallel^{1/2}} 
	\exp{ \left[ -\frac{v_\perp^2}{2R \theta_\perp} 
		-\frac{v_\parallel^2}{2R \theta_\parallel} \right]}, \hspace{0.2cm} v_\perp >0,\\
	\label{eq:II}
	\hspace{-0.36cm} \mbox{Drifted Isotropic:}\hspace{0.1cm} & f(v_\parallel,v_\perp) = \frac{C_2}{(2\pi R\theta_\parallel)^{3/2}}
	\exp{ \left[ -\frac{\left(v_\perp-\xi\right)^2}{2R \theta_\parallel} 
		-\frac{v_\parallel^2}{2R \theta_\parallel} \right]}, \hspace{0.2cm} v_\perp >0,\\
	\label{eq:III}
    \hspace{-0.36cm} \mbox{Drifted Anisotropic:}\hspace{0.1cm} & f(v_\parallel,v_\perp) = \frac{C_3}{(2\pi R)^{3/2}\theta_\perp\theta_\parallel^{1/2}}
	\exp{ \left[ -\frac{\left(v_\perp-\xi\right)^2}{2R \theta_\perp} 
		-\frac{v_\parallel^2}{2R \theta_\parallel} \right]}, \hspace{0.2cm} v_\perp >0,
	\end{align}
	\label{eq:guess}
\end{subequations}
where $C_i$ are constants which make the velocity distribution functions normalised to unity, $v_\perp$, $v_\parallel$ are the velocity components normal and parallel to the liquid-vapor interface, and $\xi$, $\theta_\perp$, $\theta_\parallel$ are the fitting free parameters. 
Note that $\xi$ cannot be identified with the mean velocity in the normal $z$-direction since the velocity distribution function is defined only for $v_\perp>0$. Likewise, $\theta_\perp$ and $\theta_\parallel$ cannot be identified with the normal and parallel temperatures. 

The values of the fitting parameters are listed in Table~\ref{tab:fits} for the lowest and highest values of the liquid bulk temperatures considered in the simulation campaign. The sum of squares of residuals, $\chi^2$, which is an indicator of the goodness of the fit, is also reported. According to $\chi^2$, the drifted anisotropic half-Maxwellian provides the best approximation, especially for the highest temperature $T_\ell/T_c=0.729$.

\begin{figure}[p!]
	\begin{subfigure}[b]{0.76\textwidth}
		\begin{center}
			\includegraphics[width=\linewidth]{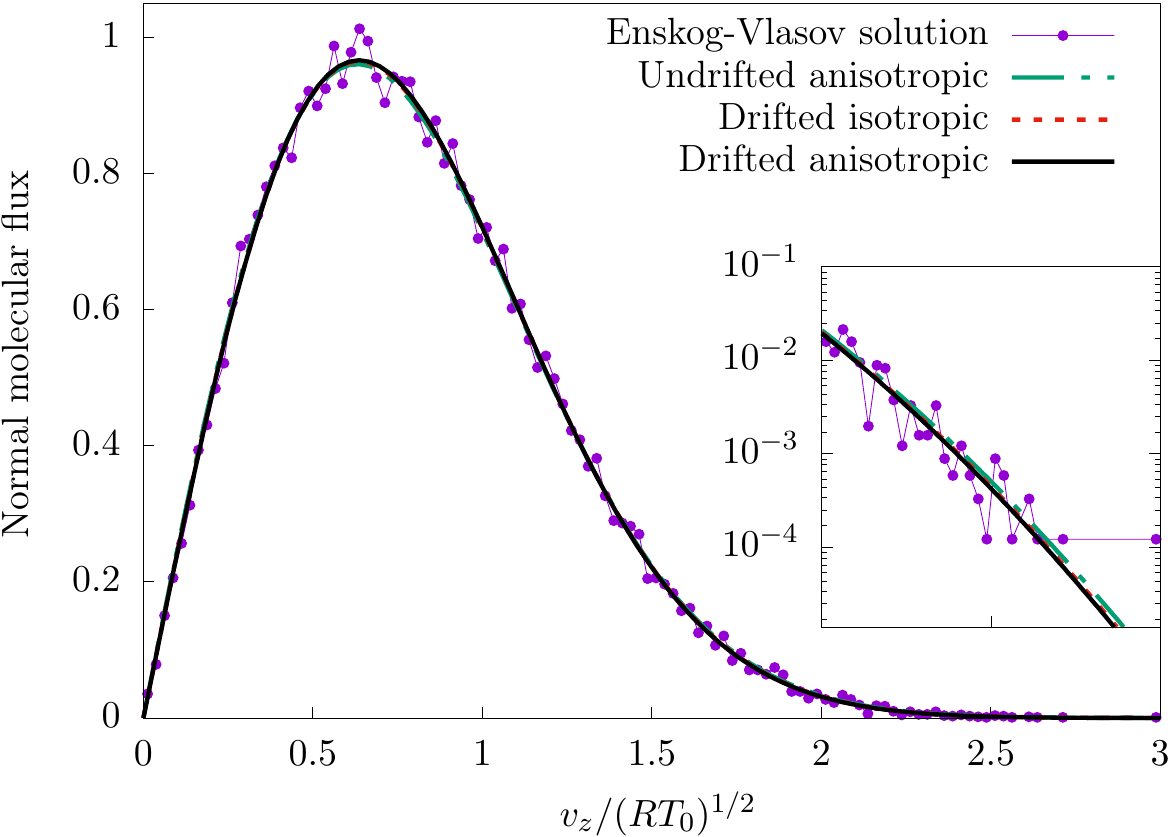}
			\subcaption{Normal molecular flux}
		\end{center} 
	\end{subfigure} 
	
	\begin{subfigure}[b]{0.5\textwidth}
		\begin{center}
			\includegraphics[width=\linewidth]{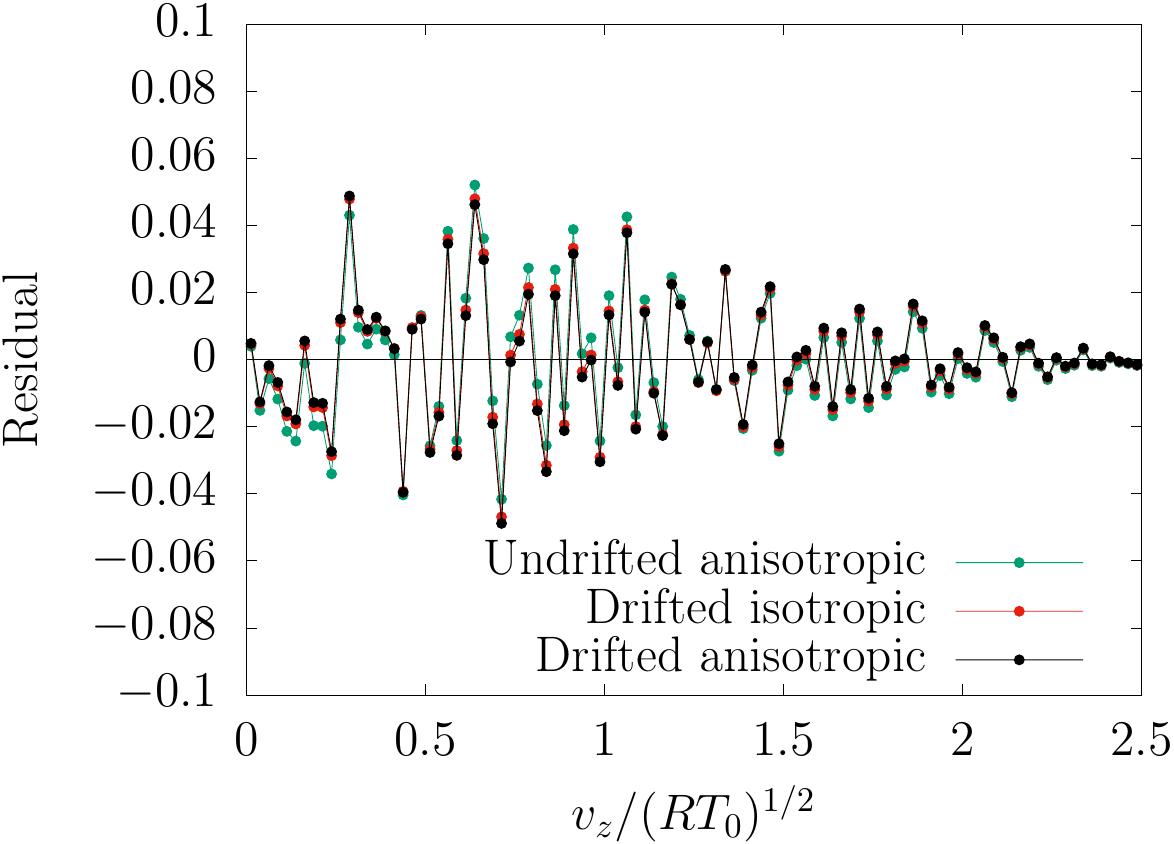}
			\subcaption{Residuals}
		\end{center}
	\end{subfigure} \hfill
	\begin{subfigure}[b]{0.395\textwidth}
		\begin{center}
			\includegraphics[width=\linewidth]{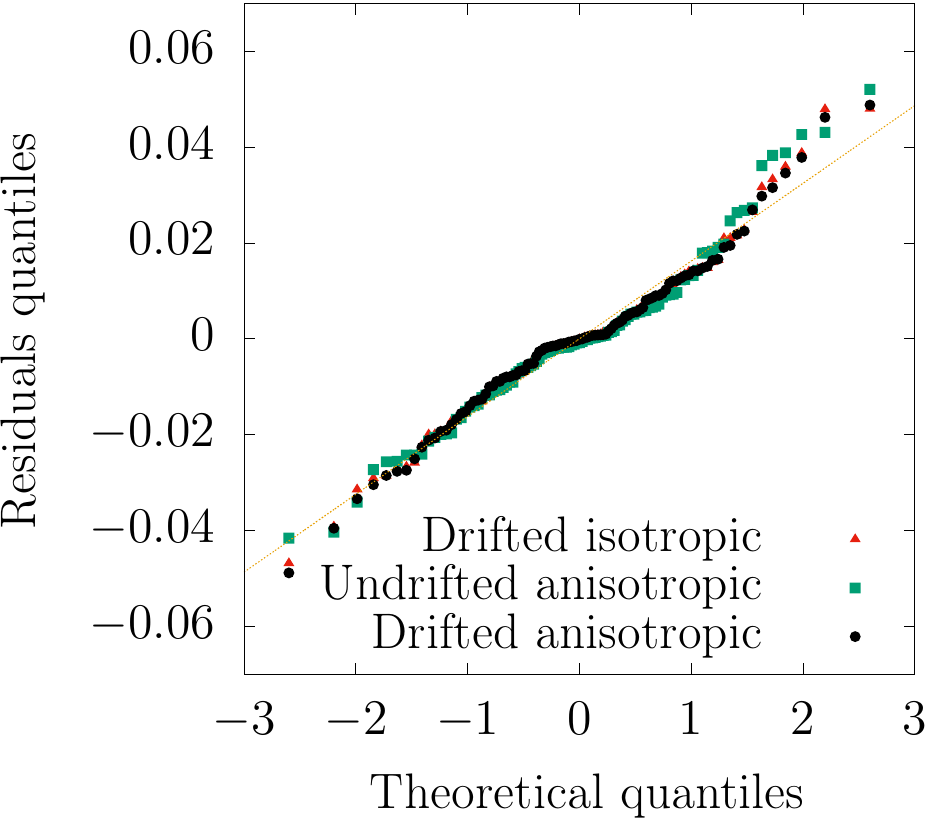}
			\subcaption{Q-Q plot}
		\end{center}
	\end{subfigure}    
	\caption{(a) Reduced normal molecular fluxes normalized to unity of evaporated molecules, (b) residual distribution, and (c) corresponding Q-Q plots of the residuals for the lowest liquid bulk temperature considered in the simulation campaign ($T_\ell/T_c=0.53$). Dashed-dotted line: Fitted undrifted anisotropic half-Maxwellian; Dashed: Fitted isotropic drifted half-Maxwellian; Solid line: Fitted drifted anisotropic half-Maxwellian. Inset: The tail of the distributions in logarithmic scale, with x axis tick values matched to the ticks of the plot.}
	\label{fig:log_plot040}
\end{figure}

\begin{figure}[p!]
	\begin{subfigure}[b]{0.76\textwidth}
		\begin{center}
			\includegraphics[width=\linewidth]{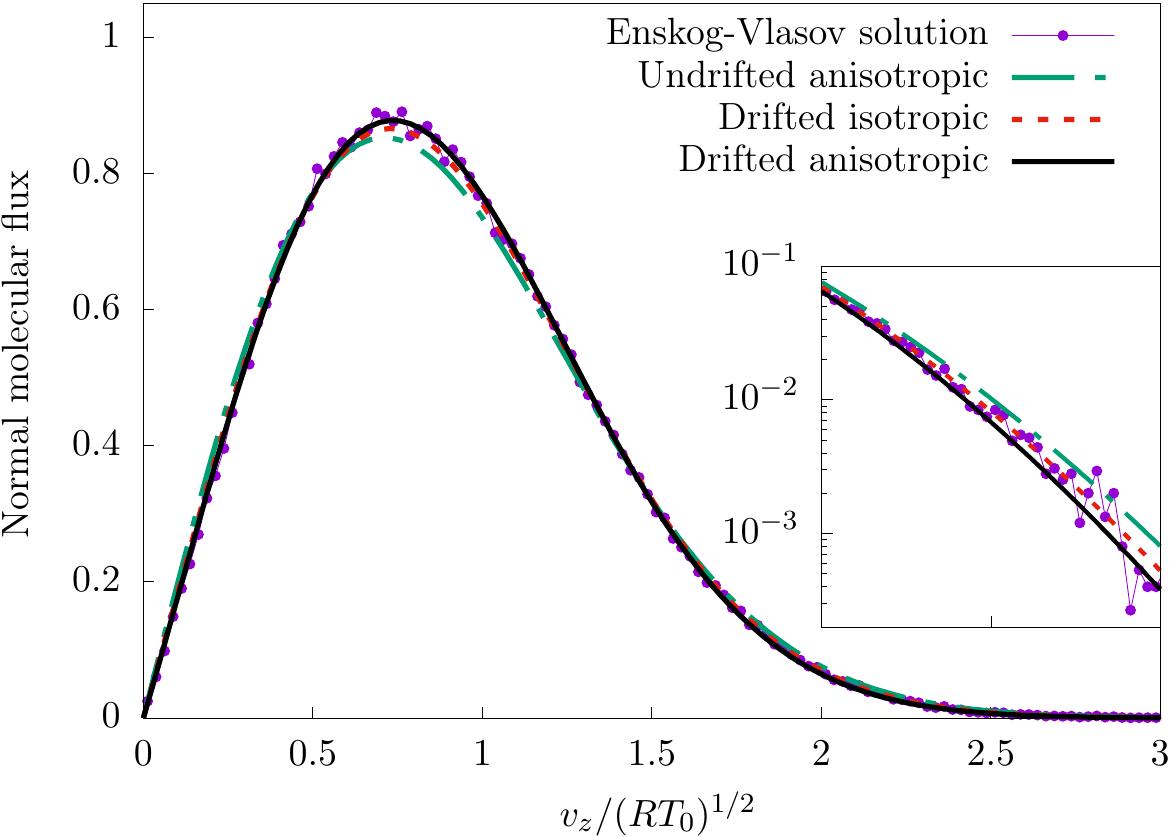}
			\subcaption{Normal molecular flux}
		\end{center} 
	\end{subfigure} \hfill
	\begin{subfigure}[b]{0.5\textwidth}
		\begin{center}
			\includegraphics[width=\linewidth]{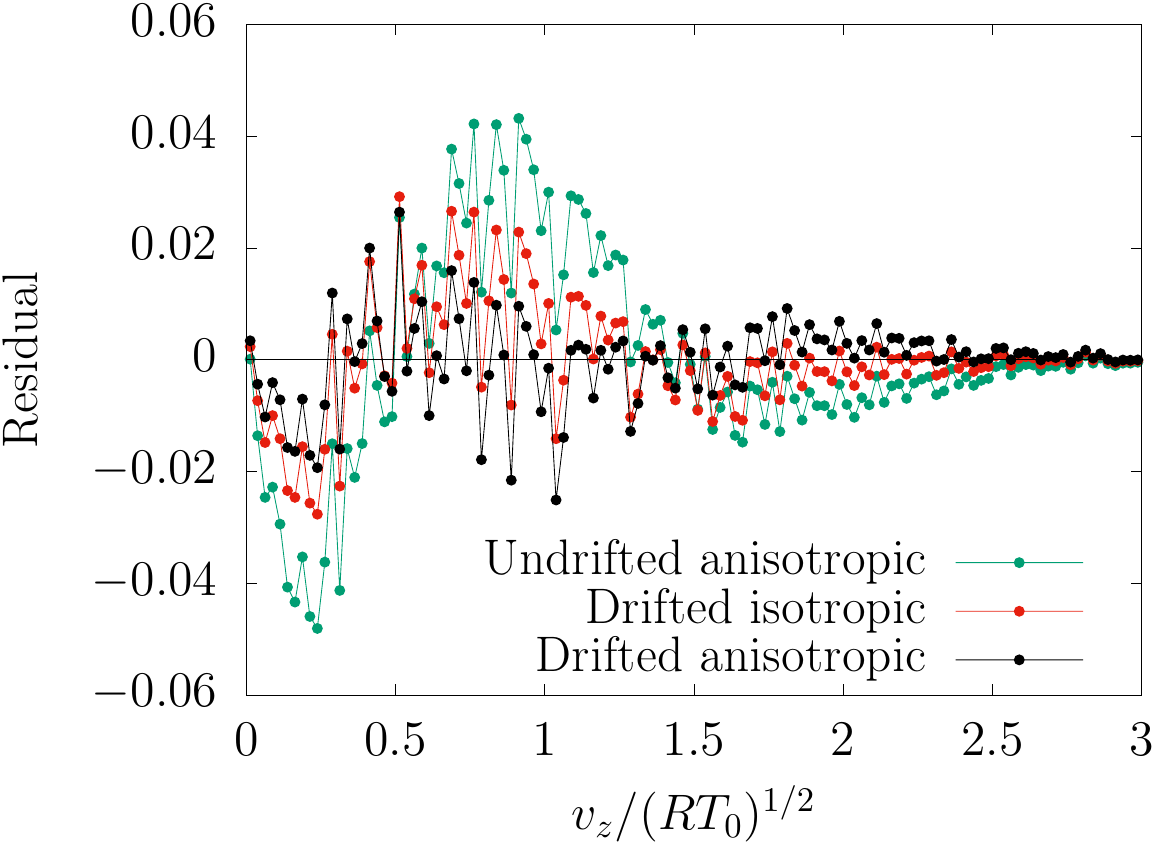}
			\subcaption{Residuals}
		\end{center}
	\end{subfigure} \hfill
	\begin{subfigure}[b]{0.4\textwidth}
		\begin{center}
			\includegraphics[width=\linewidth]{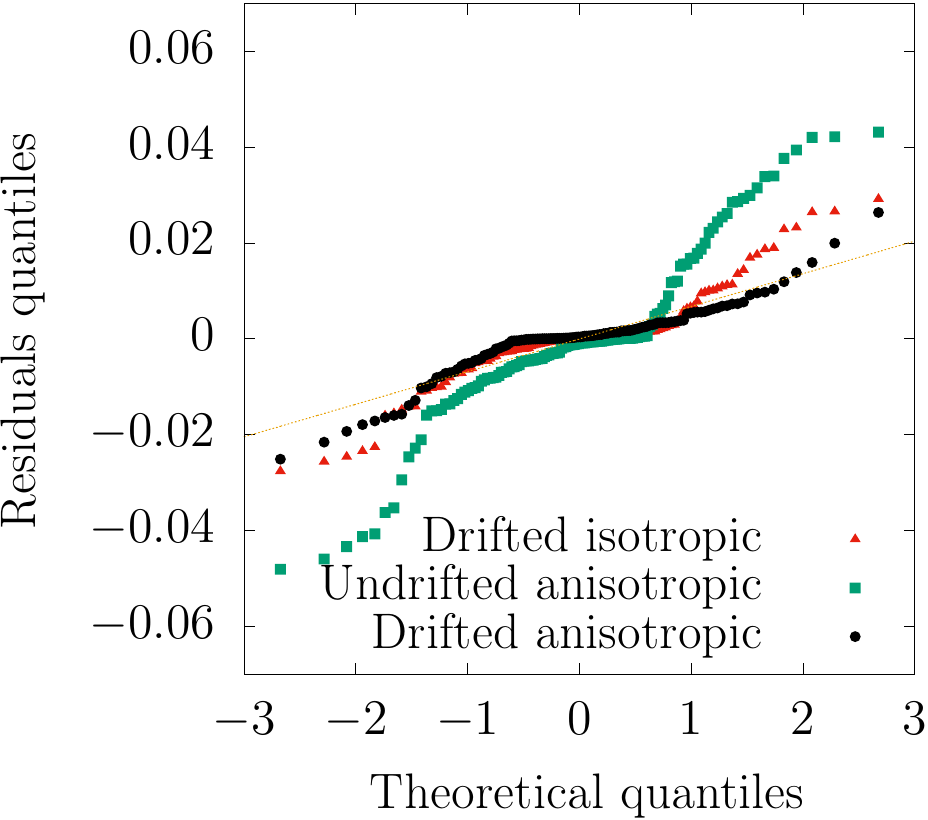}
			\subcaption{Q-Q plot}
		\end{center}
	\end{subfigure} 
	\caption{(a) Reduced normal molecular fluxes normalized to unity of evaporated molecules, (b) residual distribution, and (c) corresponding Q-Q plots of the residuals for the highest liquid bulk temperature considered in the simulation campaign ($T_\ell/T_c=0.729$). Dashed-dotted line: Fitted undrifted anisotropic half-Maxwellian; Dashed: Fitted isotropic drifted half-Maxwellian; Solid line: Fitted drifted anisotropic half-Maxwellian. Inset: The tail of the distributions in logarithmic scale, with x axis tick values matched to the ticks of the plot.}
	\label{fig:log_plot055}
\end{figure}

The direct comparison between the fitting curves and the data is shown in Figs.~\ref{fig:log_plot040} and \ref{fig:log_plot055} alongside their residuals and residuals Q-Q scatter plots which provide a more quantitative assessment of the goodness of fit. More specifically, the Q-Q scatter plot compares the quantiles of the residuals distribution functions against one another. The alignment along the bisector of first quadrant indicates that residuals are normally distributed as it should be for the ideal fitting.

In the lowest temperature case shown in Fig.~\ref{fig:log_plot040}, all the proposed distributions superimpose almost perfectly and their Q-Q scatter plots indicate a near normal distribution of residuals. By contrast, in the highest temperature case shown in Figs.~\ref{fig:log_plot055}, deviations can be observed between the different distributions, with the drifted anisotropic half-Maxwellian clearly providing the best fit. 
The direct inspection of residuals distributions, Fig.~\ref{fig:log_plot040}b and Fig.~\ref{fig:log_plot055}b, shows that the undrifted anisotropic half-Maxwellian and the drifted isotropic half-Maxwellian do not provide a good fit in the peak and tail regions, while residuals of the drifted anisotropic half-Maxwellian are randomly distributed except for the tail region where a weak pattern is visible.

None of the fitting functions perfectly matches the data but the analysis above indicates that the drifted anisotropic half-Maxwellian provides the best approximation, albeit at the cost of an extra fitting parameter. In the considered range of temperatures, using a distribution function with only one fitting parameter leads to errors in the mean velocity and temperature within one percent but errors rapidly increases if one considers higher-order moments which are more sensitive to the accuracy of the fitting in the tail region. Accordingly, for the remainder of the paper, we assume that the velocity distribution function of evaporated atoms is an anisotropic drifted half-Maxwellian.

\begin{figure}[t!]
	
	\begin{subfigure}[b]{0.49\textwidth}
		\begin{center}
			\includegraphics[width=\linewidth]{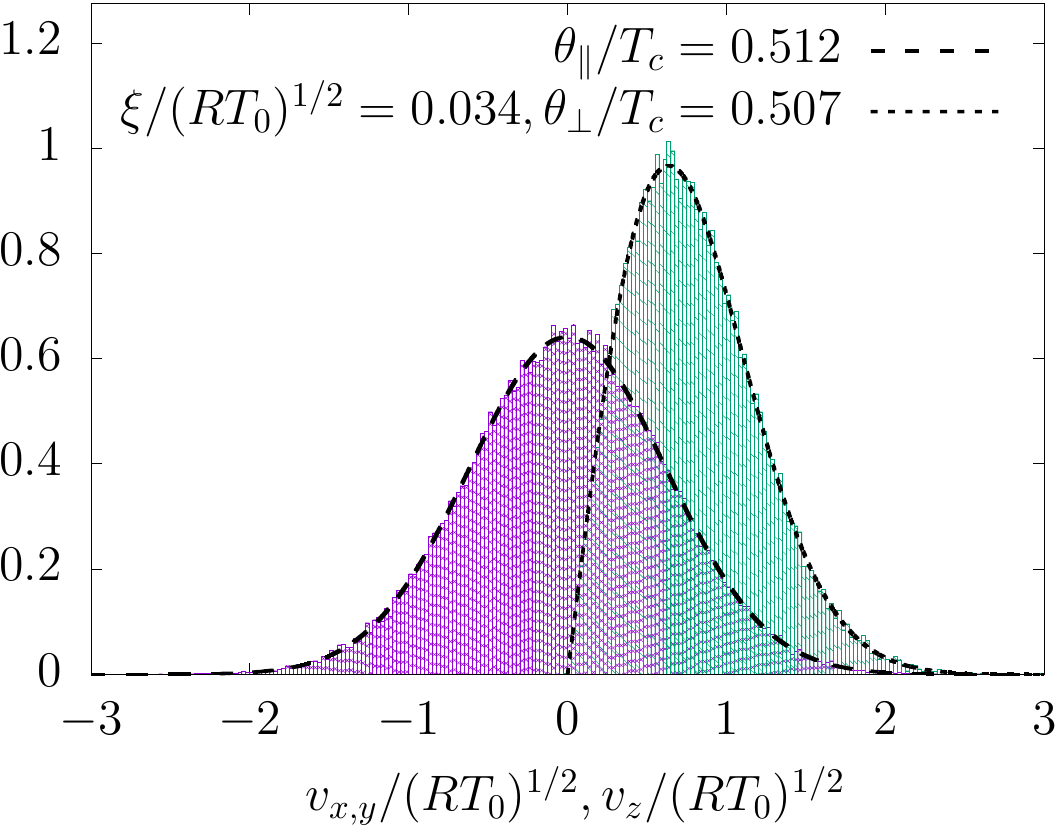}
			\subcaption{$T_\ell/T_c=0.530$}
		\end{center}
	\end{subfigure}   \hfill
	\begin{subfigure}[b]{0.49\textwidth}
		\begin{center} 			      
			\includegraphics[width=\linewidth]{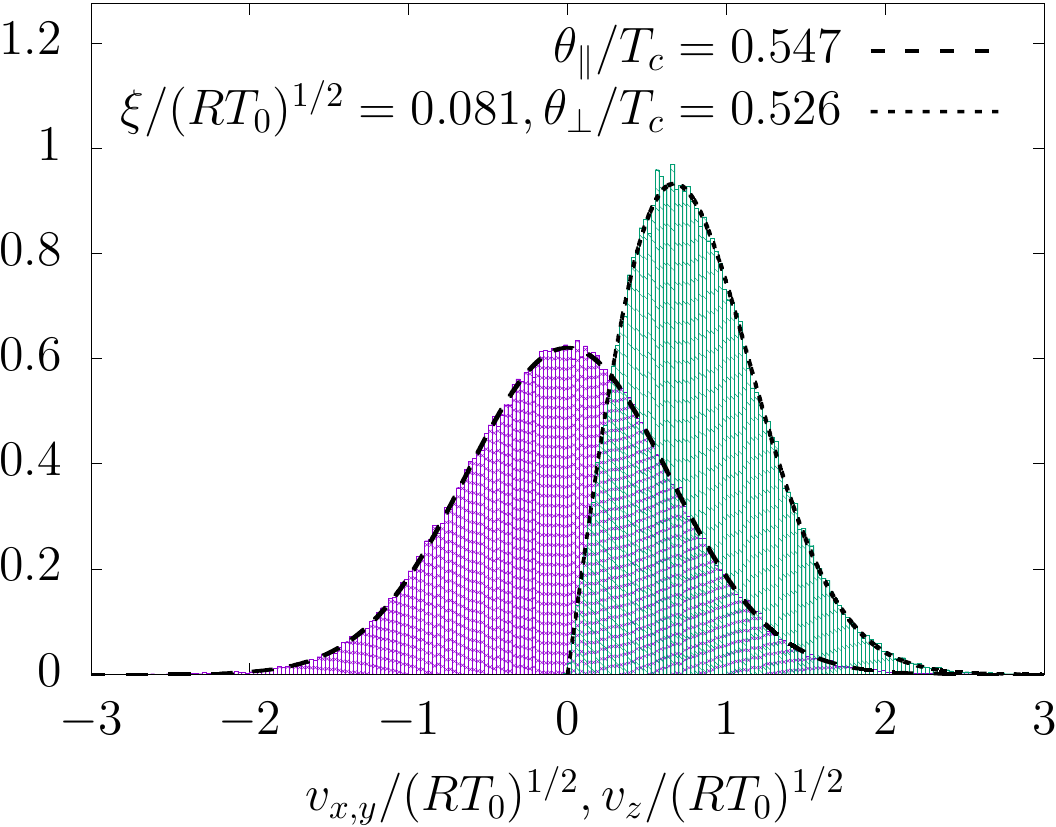}  
			\subcaption{$T_\ell/T_c=0.596$}
		\end{center}
	\end{subfigure}
	
	\vspace{0.5cm}
	
	\begin{subfigure}[b]{0.49\textwidth}
		\begin{center}
			\includegraphics[width=\linewidth]{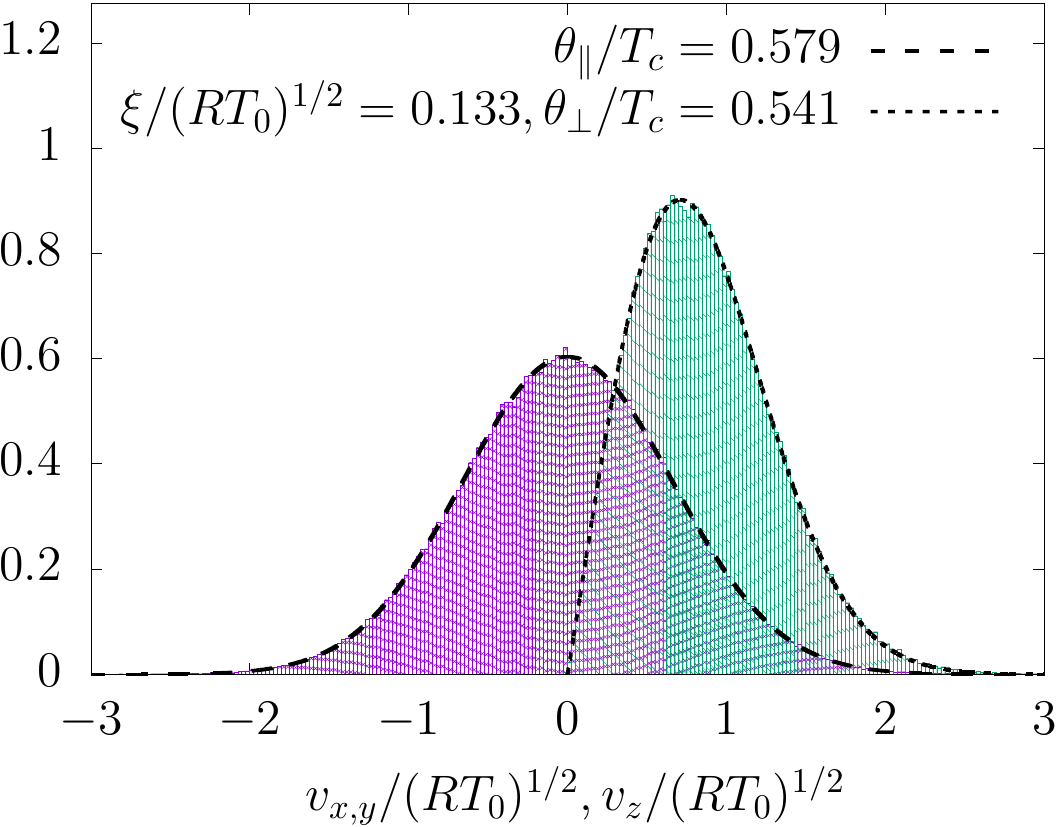}
			\subcaption{$T_\ell/T_c=0.663$}
		\end{center}
	\end{subfigure}   \hfill
	\begin{subfigure}[b]{0.49\textwidth}
		\begin{center}
			\includegraphics[width=\linewidth]{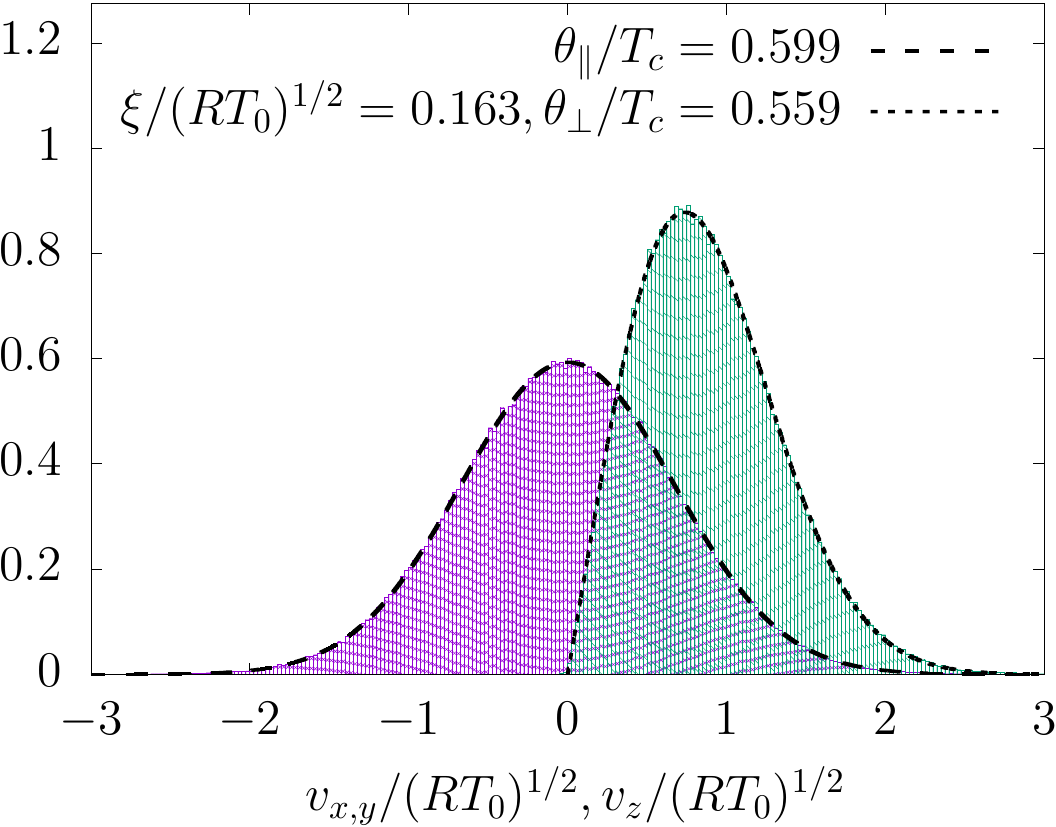}
			\subcaption{$T_\ell/T_c=0.729$}
		\end{center}		
	\end{subfigure}
	\caption{Reduced velocity distribution function and molecular flux of evaporated molecules, parallel and normal to the liquid-vapor interface, respectively, at different liquid bulk temperatures. Coloured histograms are the numerical results of the EV equation normalized to unity; solid and dashed lines are their best fits based on a drifted anisotropic half-Maxwellian with parameters $\xi$, $\theta_\perp$, and $\theta_\parallel$.}
	\label{fig:vdf}
\end{figure}

\begin{figure}[t!]
	\begin{subfigure}[b]{0.49\textwidth}
		\begin{center} 			      
			\includegraphics[width=\linewidth]{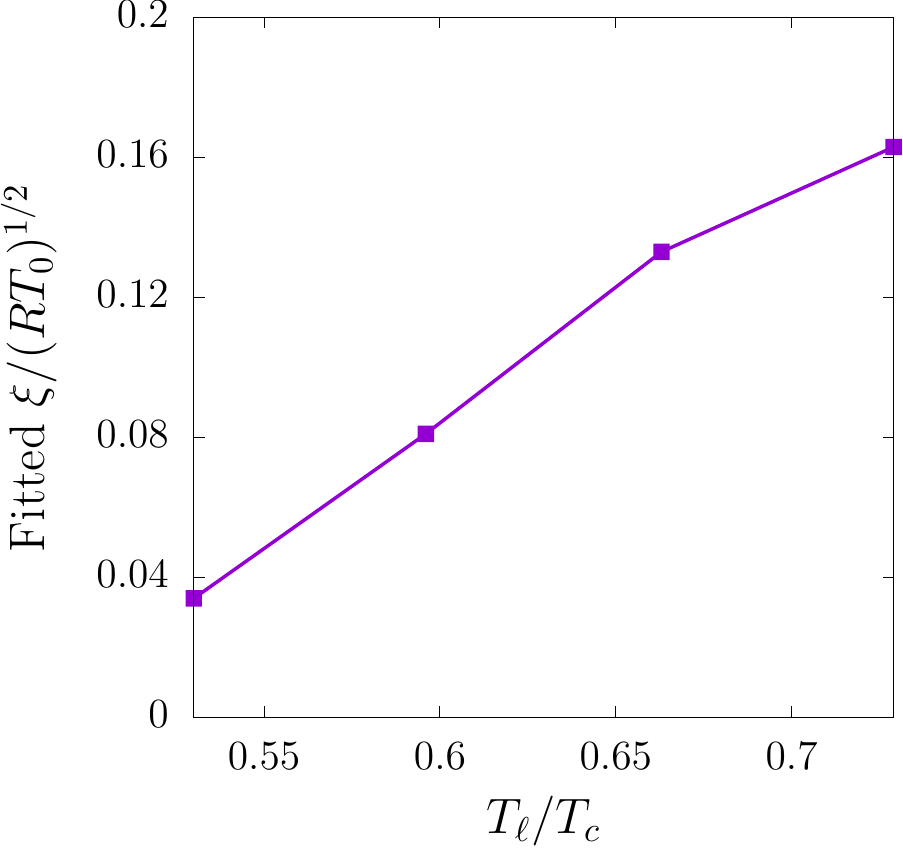}  
			\subcaption{Velocity drift.}
		\end{center}
	\end{subfigure} \hfill
	\begin{subfigure}[b]{0.49\textwidth}
		\begin{center}
			\includegraphics[width=\linewidth]{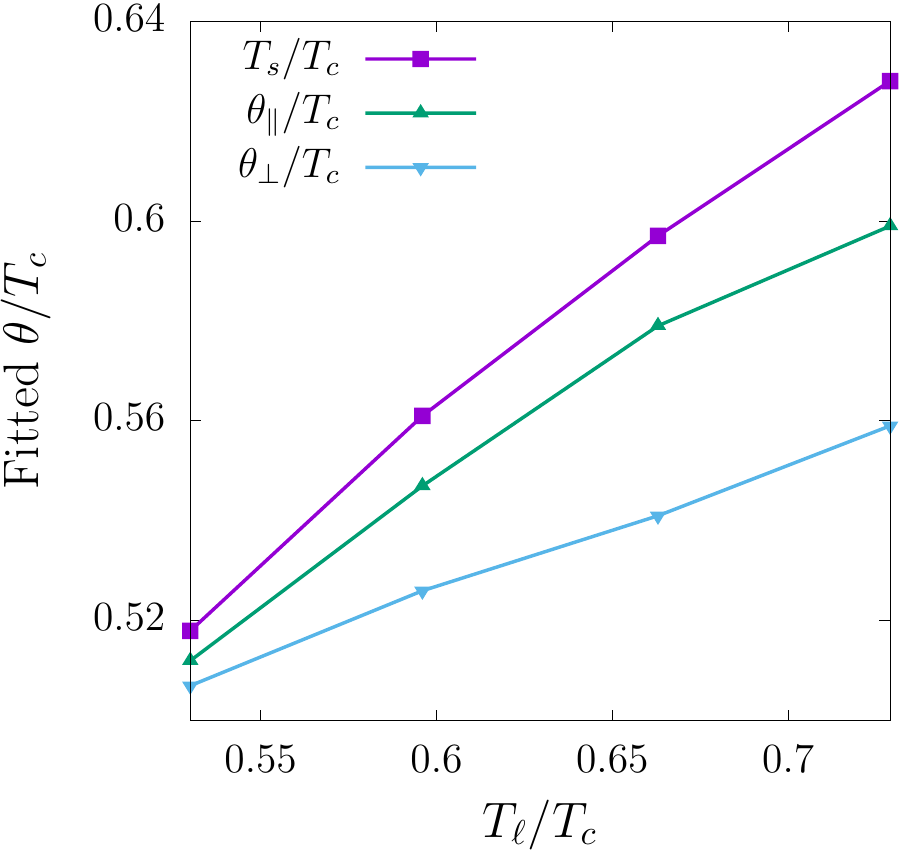}
			\subcaption{Parallel and normal temperatures.}
		\end{center}
	\end{subfigure}
    \caption{Parameters of the velocity distribution function of spontaneously evaporating atoms as a function of the liquid bulk temperature.}
    \label{fig:temp_anis}
\end{figure}

The reduced distribution function of the parallel velocity components $v_x$ and $v_y$, denoted $f_{x,y}$, and the reduced flux of the normal-velocity component $v_z$, denoted $v_zf_z$, are shown in Fig.~\ref{fig:vdf} while the values of $\xi$, $\theta_\perp$, and $\theta_\parallel$ are summarised in Table~\ref{table:temp_anis} and plotted in Fig.~\ref{fig:temp_anis}. It can be observed from Fig.~\ref{fig:temp_anis}(b) that as the liquid bulk temperature increases, the difference between the normal and parallel fitting parameters becomes larger. The drift velocity also grows monotonically with the bulk temperature, as it can be seen in Fig.~\ref{fig:temp_anis}(a).  

It is worth stressing that, the compressibility factor of the vapor is only slightly less than one in the entire range of evaporation temperatures considered in the simulation campaign (see Table~\ref{table:temp_anis}). Accordingly, the velocity drift and temperature anisotropy pointed out in this study are expected to show up even in vapors whose behaviour is only slightly non-ideal.
Note that, in the previous molecular dynamics studies~\cite{CLPG11,IYF04,MFYH04} the deviations from the undrifted isotropic half-Maxwellian were generically attributed to the collisions of atoms in the liquid-vapor interface. We will examine this argument in more depth in the next section. 

\begin{table}
\begin{center}
\begin{tabular}{ccccccc} \hline \hline
 $T_\ell/T_c$&\qquad $T_s/T_c$ &\qquad$\xi/(RT_0)^{1/2}$ &\qquad$\theta_\perp/T_c$ &\qquad$\theta_\parallel/T_c$  &\qquad$n_v a^3$ &\qquad$Z$\\  \hline 
 
0.530 &\qquad $0.518$ &\qquad 0.034 &\qquad$0.507$ &\qquad$0.512$ &\qquad 0.0008 &\qquad 0.990 \\
0.596 &\qquad$0.561$ &\qquad 0.081 &\qquad$0.526$ &\qquad$0.547$ &\qquad 0.0018 &\qquad 0.981\\
0.663 &\qquad$0.597$ &\qquad 0.133 &\qquad$0.541$ &\qquad$0.579$ &\qquad 0.0028 &\qquad 0.971\\
0.729 &\qquad$0.628$ &\qquad 0.163 &\qquad$0.559$ &\qquad$0.599$  &\qquad 0.0040 &\qquad 0.961 \\ \hline \hline
\end{tabular}
\end{center}
\caption{Separation temperature, $T_s/T_c$, velocity drift, $\xi/(RT_0)^{1/2}$, normal and parallel temperature, $\theta_\perp/T_c$ and $\theta_\parallel/T_c$, of the velocity distribution function of spontaneously evaporated atoms, number density in the vapor region, $n_v a^3$, and vapor compressibility index, $Z=p_v/(n_v k_B T_\ell)$, being $p_v$ the pressure in the vapor phase and $k_B$ the Boltzmann constant, as a function of the liquid bulk temperatures, $T_\ell/T_c$.}
\label{table:temp_anis}
\end{table}

\subsection{Velocity drift and temperature anisotropy}
\label{sec:anis}

\begin{figure}[t!]
	\begin{subfigure}[b]{0.49\textwidth}
		\begin{center} 			      
			\includegraphics[width=\linewidth]{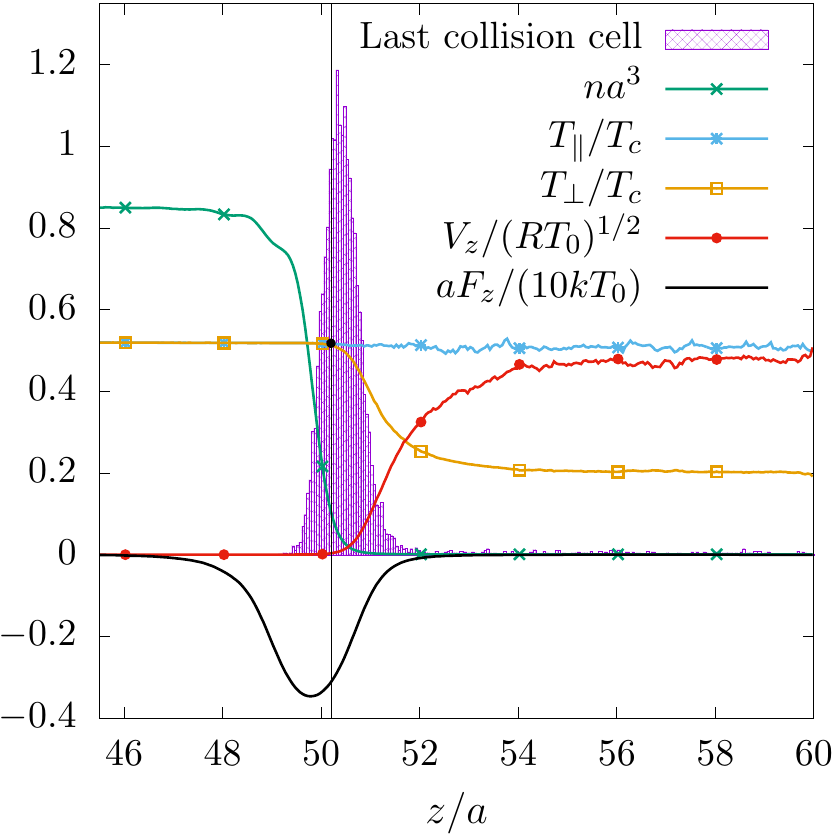}  
			\subcaption{$T_\ell/T_c=0.530$}
		\end{center}
	\end{subfigure} \hfill
	\begin{subfigure}[b]{0.49\textwidth}
		\begin{center}
			\includegraphics[width=\linewidth]{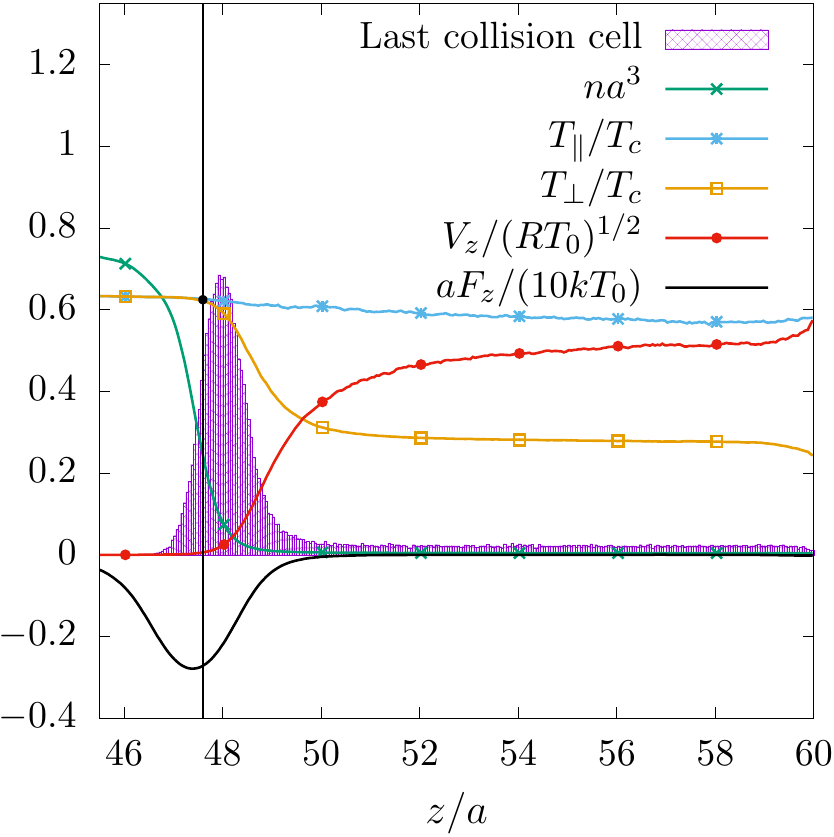}
			\subcaption{$T_\ell/T_c=0.729$}
		\end{center}
	\end{subfigure}
	\caption{Histogram of the last collision cell of evaporated atoms for the lowest and highest liquid bulk temperatures considered in the simulation campaign. The peak of the distribution is close to the separation point $z_s$ which is marked by the vertical black line.}
	\label{fig:lastcoll}
\end{figure}

\subsubsection{Numerical study}
In order to track the origin of the statistical properties of evaporated atoms, their trajectories are traced backwards, to the points where they collide for the last time. The spontaneously evaporating atoms can thus be identified as atoms that, coming from different regions of the domain, reach the absorbing surfaces in free-molecular motion under the action of the conservative mean force field.
These atoms are then divided into two groups, i.e. those whose last collision takes place before and after the separation point. Finally, the distribution functions of these two groups of atoms and the corresponding fitting parameters are evaluated.
Figure~\ref{fig:lastcoll} shows the histograms of the last collision cell of evaporated atoms for $T_\ell/T_c=0.53$ and $T_\ell/T_c=0.729$. 
As reported in Table~\ref{tab:before_after}, most of the atoms collected at the absorbing surfaces had their last collision after the separation point, i.e. they represent more than $80\%$ of the total evaporated atoms. Herein, the total number of evaporated atoms is denoted $N_e$.
Remarkably, the velocity drift and temperature anisotropy is much larger for atoms whose last collision was before $z_s$, where the local velocity distribution functions are undrifted and isotropic Maxwellians. It is thus reasonable to focus on this group of atoms to shed light on the mechanism that leads to the velocity drift and temperature anisotropy.

\begin{table}[t!]
	\begin{center}
		\begin{tabular*}{\columnwidth}{@{\extracolsep{\stretch{1}}}*{9}{c}@{}}
			\toprule
			& \multicolumn{4}{c}{Before $z_s$} & \multicolumn{4}{c}{After $z_s$} \\ \hline
			$T_\ell/T_c$  & $\xi/(RT_0)^{1/2}$ & $\theta_\perp/T_c$ & $\theta_\parallel/T_c$ & $N_{e,z<z_s}/N_e$ 
			& $\xi/(RT_0)^{1/2}$ & $\theta_\perp/T_c$ & $\theta_\parallel/T_c$ & $N_{e,z>z_s}/N_e$ \\ \hline
			0.530
			& 0.154 & 0.455 & 0.495 & 17.8\%
			& 0.036 & 0.512 & 0.518 & 82.2\% \\ 
			0.729
			& 0.344 & 0.484 & 0.567 & 11.2\%
			& 0.161 & 0.559 & 0.602 & 88.8\% \\ \hline \hline	
		\end{tabular*}
	\end{center}
	\caption{Parameters of the velocity distribution function of spontaneously evaporating atoms before and after the separation temperature for the lowest and largest liquid bulk temperatures considered.}
	\label{tab:before_after}
\end{table}

\begin{figure}[t!]
	
	\begin{subfigure}[b]{0.49\textwidth}
		\begin{center}
			\includegraphics[width=\linewidth]{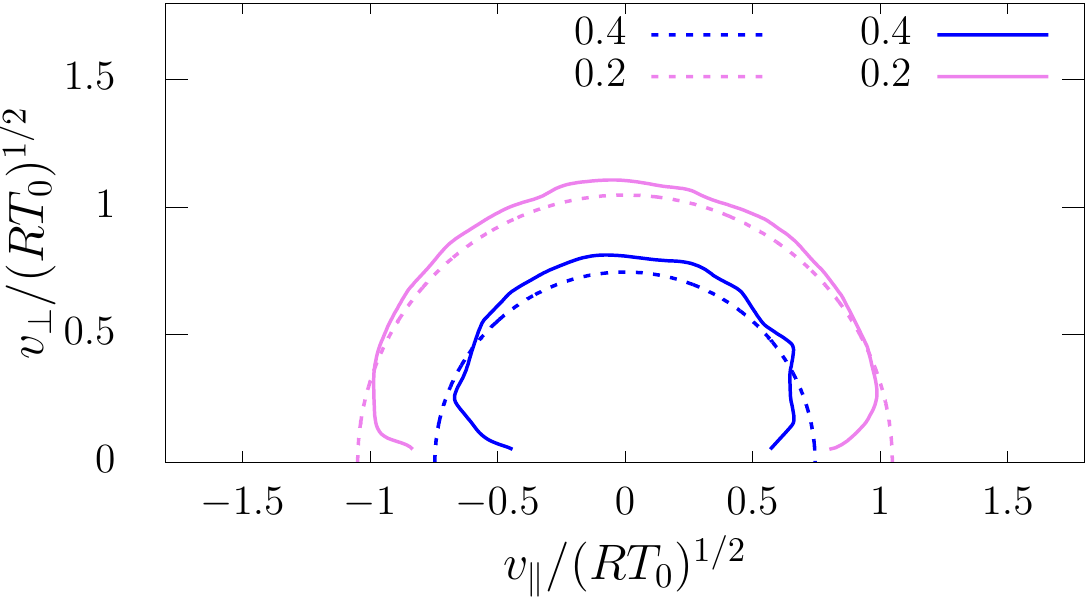}
			\subcaption{$T_\ell/T_c=0.53$}
		\end{center}
	\end{subfigure}   \hfill
	\begin{subfigure}[b]{0.49\textwidth}
		\begin{center} 			      
			\includegraphics[width=\linewidth]{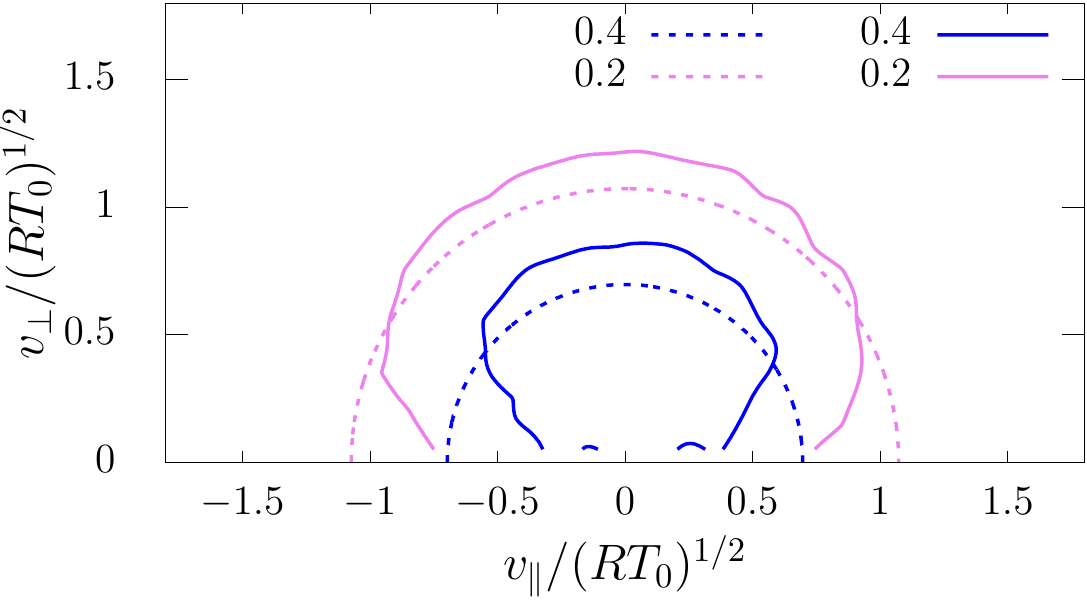}  
			\subcaption{$T_\ell/T_c=0.729$}
		\end{center}
	\end{subfigure}
\caption{Isocontours of the distribution functions of evaporated atoms (dashed lines) and `potentially' evaporating atoms (solid lines). Atoms originate from locations before the separation point.}
\label{fig:fz_contours}	
\end{figure}

In Fig.~\ref{fig:fz_contours} we plot the isocontours of the velocity distribution functions of the evaporated atoms originating before the separation point (group A) and of the `potentially' evaporating atoms (group B). This latter group comprises all the atoms coming from the same locations of the ones of group A under the assumption that they can reach the absorbing surface without suffering any backscattering due to collisions in the interface region. 
Both velocity distribution functions are normalized to unity.
The velocity distribution function of the group B is an undrifted isotropic half-Maxwellian with the weighted average temperature of the region where they originate.  
This is not unexpected, since, as also proved in Section~\ref{sec:toy_model}, an undrifted Maxwellian velocity distribution function of atoms moving in a conservative force field remains Maxwellian with the same temperature. 
The comparison clearly shows that the evaporated atoms have predominantly higher normal velocities and lower parallel velocities. This result can be understood as the result of the interplay between normal and parallel velocities due to collisions in the liquid-vapor interface. Indeed, the higher is the normal-velocity component, the less likely an atom suffers a collision in the liquid-vapor interface because the lower is the time spent in that region. However, at the same time, the larger is the speed, the greater becomes the probability of collisions (see also Eq.~\eqref{eq:colf}). This interplay also explains why atoms coming from the region before the separation point mostly contribute to the velocity drift and temperature anisotropy. Indeed, these atoms must travel a larger distance before reaching the absorbing surface and are thus more likely to be back-scattered by collisions. Likewise, the liquid-vapor interface gets wider as the liquid bulk temperature increases and, therefore, the effect of preferential evaporation of atoms with large normal and lower parallel velocity components is enhanced.

\subsubsection{An instructive model}
\label{sec:toy_model}

\begin{table}[t!]
	\begin{center}
		\begin{tabular*}{0.6\columnwidth}{@{\extracolsep{\stretch{1}}}*{4}{cccc}@{}}
			\toprule
		     $T_\ell/T_c$ & $\bar{d}/a$ & $\bar{n}a^3$ & $\bar{T}/T_c$ & $\Delta\mathcal{U}/(k_B T_s)$ \\ \hline
		     0.530 & 9.8 & 0.00157 & 0.44 & 0.01686 \\
		     0.729 & 12.4 & 0.00554 & 0.497 & 0.04936 \\ \hline \hline
		\end{tabular*}
	\end{center}
	\caption{Numerical values of the parameters which enter in the model for the lowest and highest liquid bulk temperatures considered in the simulation campaign
}
    \label{tab:toy}
\end{table}

\begin{figure}[t!]
	\begin{subfigure}[b]{0.49\textwidth}
		\begin{center}
			\includegraphics[width=\linewidth]{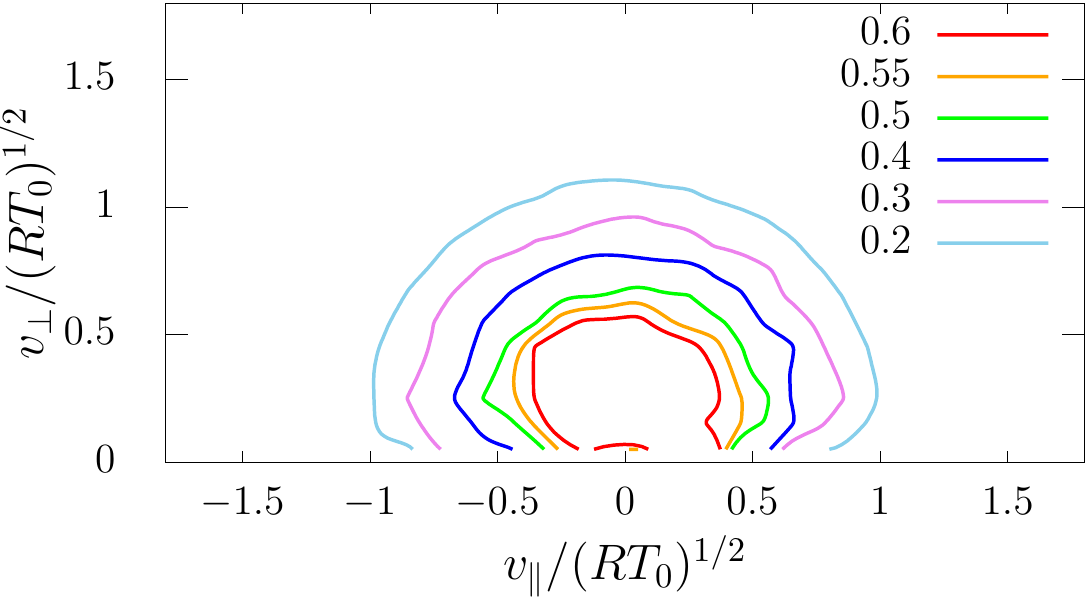}
			\subcaption{EV solution, $T_\ell/T_c=0.53$}
		\end{center}
	\end{subfigure}   \hfill
	\begin{subfigure}[b]{0.49\textwidth}
		\begin{center} 			      
			\includegraphics[width=\linewidth]{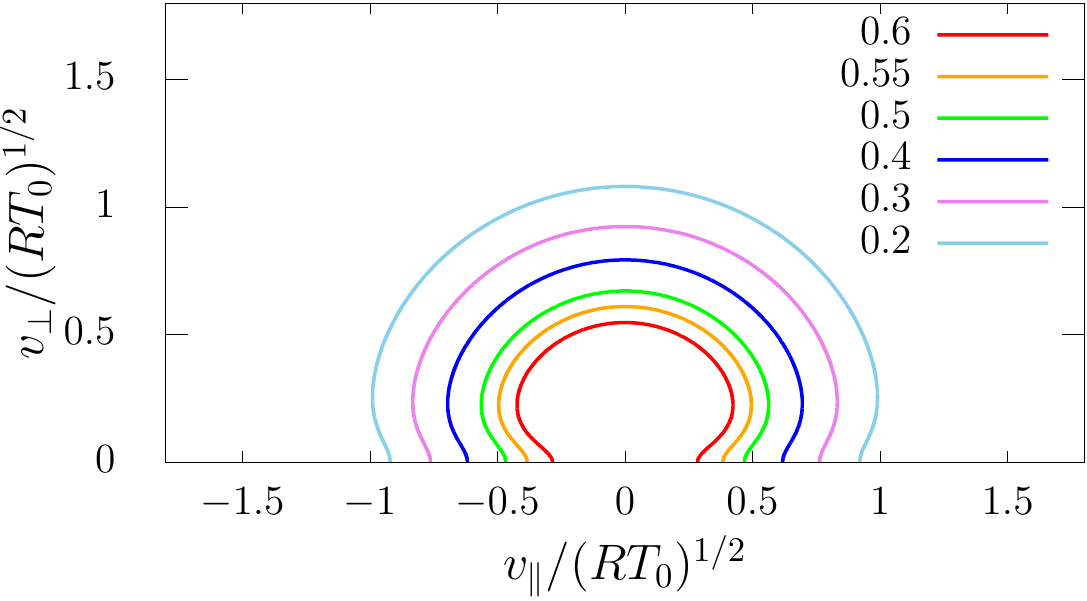}  
			\subcaption{Model prediction, $T_\ell/T_c=0.53$}
		\end{center}
	\end{subfigure}
	
	\vspace{0.5cm}
	
	\begin{subfigure}[b]{0.49\textwidth}
		\begin{center}
			\includegraphics[width=\linewidth]{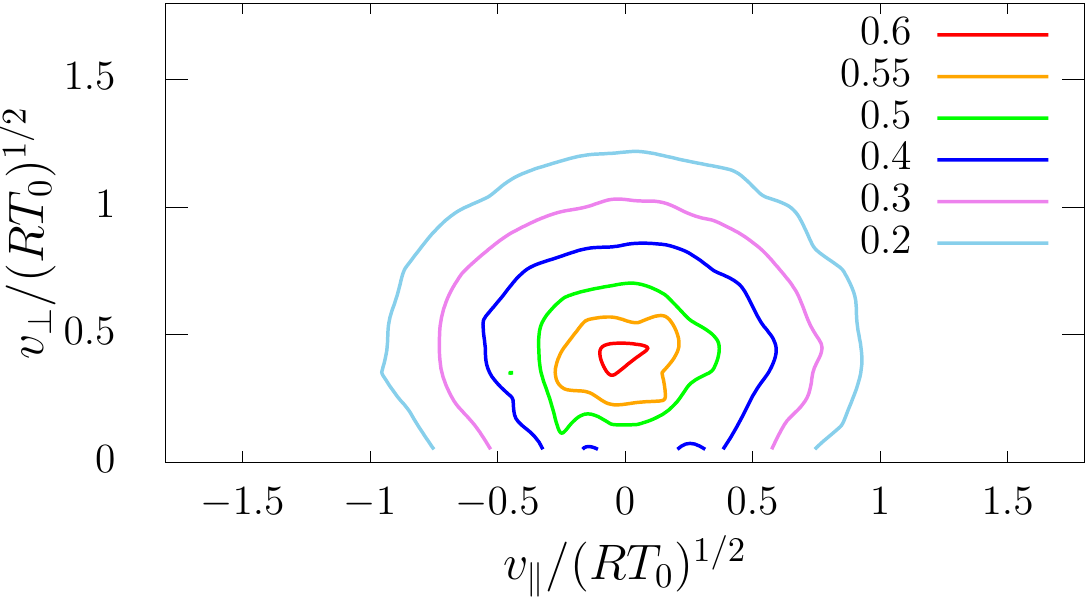}
			\subcaption{EV solution, $T_\ell/T_c=0.729$}
		\end{center}
	\end{subfigure}   \hfill
	\begin{subfigure}[b]{0.49\textwidth}
		\begin{center}
			\includegraphics[width=\linewidth]{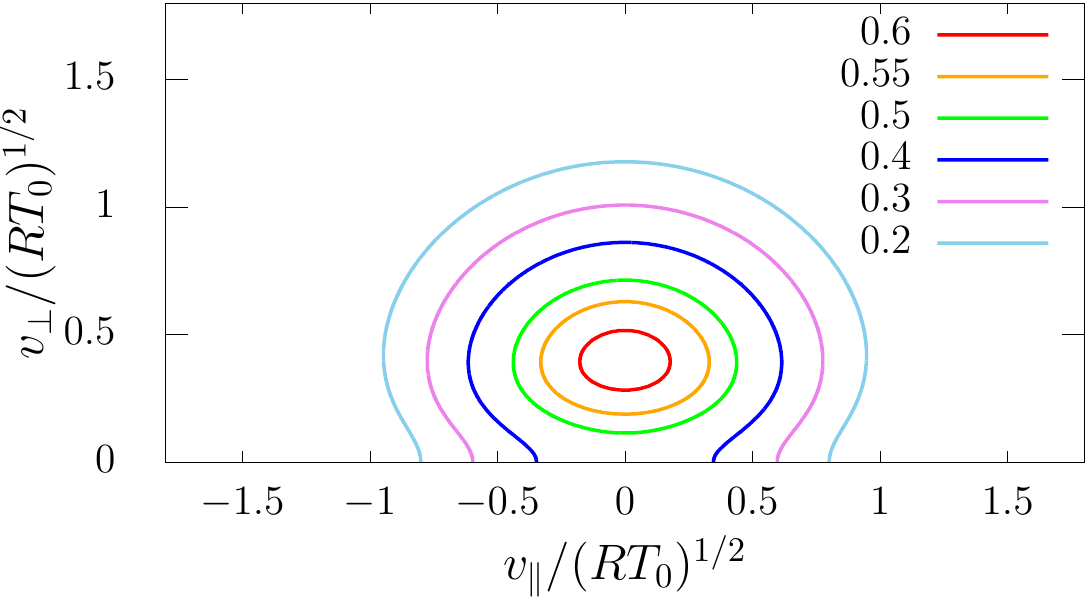}
			\subcaption{Model prediction, $T_\ell/T_c=0.729$}
		\end{center}		
	\end{subfigure}
	\caption{Isocontours of distribution functions of evaporated particles for two evaporation temperatures.}
	\label{fig:toy_contours}
\end{figure}

The results presented in the previous section suggest that the velocity drift and the temperature anisotropy of spontaneously evaporating atoms is a consequence of the collisions in the liquid-vapor interface which filter out atoms with a lower normal-velocity component. 

In order to support this conclusion, here we approximately evaluate the distribution function of atoms which, being initially at the separation point, cross the liquid-vapor interface, and are collected at the absorbing surface placed at a distance $\bar{d}$. For simplicity, it is assumed that atoms move in a region of constant density, $\bar{n}$, and temperature, $\bar{T}$.
At the separation point, atoms are supposed to be distributed according to an isotropic Maxwellian with zero drift and temperature $T_s$:
\begin{equation}
\label{eq:f_s}
f_s\left(v_{\parallel},v_\perp \right) = \frac{n_s}{\left(2\pi R T_s\right)^{3/2}} \exp\left( -\frac{v_\parallel^2+v_\perp^2}{2RT_s} \right).
\end{equation}
Only a fraction of these atoms can reach the absorbing surface, namely the ones which (i) have a speed sufficiently large to overcome the potential barrier of the mean force field which pushes atoms towards the liquid phase and (ii) are not backscattered due to the collisions with other atoms. Note that:
\begin{itemize}
\item[(i)] The conservation of mechanical energy implies:
\begin{equation}
\label{eq:energy}
\frac{1}{2} m v_s^2 + {\mathcal U}_s = \frac{1}{2} m v_e^2 + {\mathcal U}_e \, \Longrightarrow \, v_{\perp,s}^2 = v_{\perp,e}^2 +v_{\perp,min}^2, \hspace{0.5cm} v_{\perp,min}=\left(2 \frac{\Delta \mathcal{U}}{m}\right)^{1/2},
\end{equation} 
where $(v_s,{\mathcal U}_s)$ and $(v_e, {\mathcal U}_e)$ are the atom's velocity and the potential energy of the mean force field at the separation point and at the absorbing surface, respectively, and $\Delta \mathcal{U}=\mathcal{U}_e-\mathcal{U}_s$.
In Eq.~\eqref{eq:energy}, it has been used that $v_{\parallel,s}=v_{\parallel,e}$ since the mean force field acts along the $z$-direction. It is plain that only atoms whose $z$-component of the velocity is larger than $v_{\perp,min}$ can reach the absorbing surface, all the others turn back, under the action of the force field.
\item[(ii)] The number of collisions experienced by an atom in the time interval $\Delta t$ can be assumed to follow a Poisson distribution:
\begin{subequations}
\begin{equation}
Pr(\mbox{$N$ Collisions in the time interval $\Delta t$})=\frac{(\bar{\nu} \Delta t)^Ne^{-\bar{\nu} \Delta t}}{N!},
\end{equation}
where the collision rate $\bar{\nu}$ can be estimated as:
\begin{equation}
\bar{\nu}(v)= \chi(\bar{n}) \frac{\pi a^2 \bar{n}}{\sqrt{\pi}\beta}
\left[e^{-\beta^2 v^2}+\left(2 \beta v+\frac{1}{\beta v}\right)\frac{\pi}{2}\erf(\beta v)\right],
\label{eq:colf}
\end{equation}
where $\beta^2=1/(2R\bar{T})$~\cite{K38}.
Accordingly, the probability that an atom reaches the absorbing surface without suffering any collision is:
\begin{equation}
Pr(\mbox{$0$ Collisions in the time interval $\Delta t$})=e^{-\bar{\nu}(v) \Delta t}=e^{-\bar{\nu}(v)\frac{\bar{d}}{v_\perp}},
\end{equation}
\end{subequations}
where it has been used that the time needed to cross the liquid-vapor interface is $\Delta t=\bar{d}/v_\perp$.
\end{itemize}
By adopting cylindrical coordinates and using (i)-(ii), the net mass flux of atoms across the liquid-vapor interface can be written as:
\begin{equation}
	\label{eq:balance_interface}
	\int_{0}^{+\infty} dv_{\parallel} \, v_\parallel  \int_{v_{\perp,min}}^{+\infty} dv_\perp \, f_s\left(v_{\parallel},v_\perp \right) v_\perp =
	\int_{0}^{+\infty} d\tilde{v}_{\parallel} \, \tilde{v}_\parallel \int_{0}^{+\infty} d\tilde{v}_\perp f_e \left(\tilde{v}_{\parallel},\tilde{v}_\perp \right) \tilde{v}_\perp,
\end{equation}
where $f_e$ is given by Eq.~\eqref{eq:f_s} and $f_e$ is the unknown distribution function of evaporated atoms, i.e. atoms collected at the absorbing surface. 
Note that in Eq.~\eqref{eq:balance_interface}, the integral over the azimuthal angle cancels out due to the symmetry of the system. 
The first integral on the right hand side of Eq.~\eqref{eq:balance_interface} can be simplified by  making the change of variables 
$\left( v_\perp, v_\parallel\right) \rightarrow \left( (\tilde{v}_\perp^2+\tilde{v}_{\perp,min}^2)^{1/2}, \tilde{v}_\parallel\right)$ suggested by the conservation of mechanical energy, Eq.~\eqref{eq:energy}. The distribution function of atoms collected at the absorbing surface can thus be readily obtained:  
\begin{multline}
\label{eq:f_p}
f_e\left(\tilde{v}_{\parallel},\tilde{v}_\perp \right)=
 \frac{n_s}{\left(2\pi R T_s\right)^{3/2}} 
  \exp\left( -\frac{\tilde{v}_{\perp,min}^2}{2RT_s} \right) \\
 \exp\left( -\frac{\tilde{v}_\parallel^2+\tilde{v}_\perp^2}{2RT_s} \right)  
 \exp\left[ -\frac{\bar{\nu}\left( (\tilde{v}^2+v^2_{\perp,min})^{1/2}\right)
 	\bar{d}}{\left( \tilde{v}_{\perp}^2+v_{\perp,min}^2 \right)^{1/2}} \right].
\end{multline}
Note that if one disregards collisions, $\bar{\nu}=0$, Eq.~\eqref{eq:f_p} simplifies to a half-Maxwellian with the parameter density given by $n_s$ reduced by the Boltzmann factor $\exp{[-\Delta \mathcal{U}/(k_B T_s)]}$.  


The dimensionless values of the width of the liquid-vapor interface, $\bar{d}/a$, the mean density, $\bar{n}a^3$, and temperature, $\bar{T}/T_c$, of the liquid-vapor interface are estimated from the simulation results presented in the previous sections. The potential jump $\Delta\mathcal{U}/(k_B T_s)$ is the one that occurs at the edge of a slab having length $\bar{d}/a$ and density $\bar{n}a^3$ placed next to vacuum. The numerical values of all these parameters are listed in Table~\ref{tab:toy}.
Figure~\ref{fig:toy_contours} shows the comparison between the normalised velocity distribution function of evaporated atoms obtained by numerically solving the Enskog-Vlasov equation, and the predictions given by Eq.~\eqref{eq:f_p}. The very good qualitative agreement strongly suggests that the collisions in the liquid-vapor interface may be responsible for deviations from the isotropic half-Maxwellian.

\section{Conclusions}
\label{sec:conclusions}
The Enskog-Vlasov (EV) equation has been used to study the one-dimensional steady evaporation of a monatomic liquid into near vacuum conditions. 
The main aim has been to elucidate the statistical features of atoms spontaneously emitted by the liquid bulk. This is a key step in formulating kinetic boundary conditions at the liquid-vapor interface. 

The mean-field kinetic theory approach is used in this study since it is by far less computationally demanding than molecular dynamics simulations and permits one to get results with the required high level of accuracy.

The velocity distribution function of spontaneously evaporating atoms are commonly assumed to be an half-Maxwellian at the temperature of the liquid-vapor interface. 
By contrast, the main results of this work show that:
\begin{itemize}
	\item Evaporated atoms are distributed according to a drifted anisotropic half-Maxwellian. Deviations from the isotropic half-Maxwellian become more pronounced as the liquid bulk temperature increases.
	\item The velocity drift and the temperature anisotropy are the results of collisions in the liquid-vapor interface region which preferentially backscatter atoms with a lower normal-velocity component. 
\end{itemize}
Note that previous studies based on molecular dynamics simulations have already pointed out that the distribution function of evaporating atoms deviate from an isotropic half-Maxwellian but there hasn't been unanimous agreement on the functional form, i.e. an anisotropic half-Maxwellian~\cite{TTM99,IYF04} or a drifted half-Maxwellian~\cite{MFYH04}. Furthermore, deviations were noticed only for high evaporation temperatures and, accordingly, attributed to the non-ideal vapor behaviour.
 
The novelty of the present work is thus threefold. 
First, it establishes that the drifted anisotropic half-Maxwellian provides the best fitting out of the considered distributions. 
Second, it shows that deviations from the half-Maxwellian occur even when the vapor is only slightly non-ideal, namely in conditions where, in principle, the usual kinetic-theory treatment of the evaporation process can still be used~\cite{FBG19}.
Third, it brings evidence, through accurate numerical results and a simple model, that collisions in the interface may explain the velocity drift and the temperature anisotropy.   

It is worth stressing that the results above are expected to be valid even in presence of net condensation since it is commonly assumed that the evaporation flux only depends on the state of the fluid in the liquid phase. However, a more detailed study would be needed to assess to what extent the vapor dynamics affects the structure of the liquid-vapor interface and, in turn, the evaporation mass flux.

This work paves the way to a reformulation of boundary conditions usually adopted at the liquid-vapor interface in kinetic theory studies of evaporation/condensation processes. In this respect, an interesting research perspective consists in determining the dependence of the drift-velocity and temperature anisotropy on the properties of the liquid bulk by means of the balance equations of mass, momentum, and energy derived from the EV equation.


\begin{acknowledgments}
    This work has been financially supported in the UK by EPSRC grants (EP/N016602/1, EP/R007438/1, EP/S029966/1 \& EP/P031684/1) and the Leverhulme Trust.
\end{acknowledgments}

\bibliography{bibliography}

\end{document}